\documentclass[superscriptaddress,preprintnumbers,amssymb]{revtex4} 
\usepackage{graphicx}
\usepackage{epsfig}
\usepackage{amsmath}
\usepackage{amsfonts}
\usepackage{amssymb}

\usepackage{braket}
\usepackage{cancel}
\usepackage{pstricks}
\usepackage{color}

\def\lsim{\:\raisebox{-0.5ex}{$\stackrel{\textstyle<}{\sim}$}\:}
\def\gsim{\:\raisebox{-0.5ex}{$\stackrel{\textstyle>}{\sim}$}\:}

\usepackage{epstopdf}

\begin{document}
{\tiny }\global\long\def\bra#1{\Bra{#1}}
{\tiny }\global\long\def\ket#1{\Ket{#1}}
{\tiny }\global\long\def\set#1{\Set{#1}}
{\tiny }\global\long\def\braket#1{\Braket{#1}}
{\tiny }\global\long\def\norm#1{\left\Vert #1\right\Vert }
{\tiny }\global\long\def\rmto#1#2{\cancelto{#2}{#1}}
{\tiny }\global\long\def\rmpart#1{\cancel{#1}}
{\tiny \par}

\title {Chiral heavy fermions in a two Higgs doublet model: 750 GeV resonance or not}
\author{Shaouly Bar-Shalom}
\email{shaouly@physics.technion.ac.il}
\affiliation{Physics Department, Technion-Institute of Technology, Haifa 32000, Israel}
\author{Amarjit Soni}
\email{adlersoni@gmail.com}
\affiliation{Physics Department, Brookhaven National Laboratory, Upton, NY 11973, USA}

\date{\today}

\begin{abstract}
We revisit models where a heavy chiral 4th generation doublet of fermions
is embedded in a class of two Higgs doublets models (2HDM)
with a discrete $Z_2$ symmetry, which couples the ``heavy" scalar
doublet only to the 4th generation fermions and the
``light" one to the Standard Model (SM) fermions -
the so-called 4G2HDM introduced by us several years ago.
We study the constraints imposed on the 4G2HDM from direct searches of
heavy fermions, from precision electroweak data (PEWD)
and from the measured production and decay signals of the
125 GeV scalar, which in the 4G2HDM corresponds to the lightest CP-even scalar $h$.
We then show that the recently reported excess in the $\gamma\gamma$ spectrum
around 750 GeV can be accommodated by the heavy CP-even scalar
of the 4G2HDM, $H$, resulting in a unique choice of parameter space:
 negligible mixing ($\sin\alpha \lsim {\cal O}(10^{-3})$)
between the two CP-even scalars $h,H$ and heavy 4th generation quark and lepton masses
$m_{t^\prime},m_{b^\prime} \lsim 400~{\rm GeV}$
and $m_{\nu^\prime},m_{\tau^\prime} \gsim 900~{\rm GeV}$, respectively.
Whether or not the 750 GeV $\gamma \gamma$ resonance
is confirmed, interesting phenomenology emerges
in $q^\prime$ - Higgs systems ($q^\prime=t^\prime,b^\prime$),
that can be searched for at the LHC.
For example, the heavy scalar states of the model, $S=H,A,H^+$, may have
$BR(S \to \bar q^\prime q^\prime) \sim {\cal O}(1)$, giving rise
to observable $\bar q^\prime q^\prime$ signals on resonance, followed
by the flavor changing $q^\prime$ decays $t^\prime \to uh$ ($u=u,c$)
and/or $b^\prime \to dh$ ($d=d,s,b$). This leads
to rather distinct signatures, with or without charged leptons,
of the form $\bar q^\prime q^\prime \to (nj +  mb + \ell W)_S$ ($j$ and $b$ being light and b-quark jets, respectively), with
$n+m+\ell =6-8$ and unique kinematic features.
These high jet-multiplicity signals appear to be very challenging and may need new search strategies for detection of
such heavy chiral quarks.
It is  also shown that the flavor structure of the 4G2HDM
can easily accommodate the interesting recent indications of a percent-level branching ratio
in the lepton-flavor-violating (LFV) decay $h \to \tau \mu$ of the 125 GeV Higgs, if it is
experimentally confirmed.
\end{abstract}


\maketitle

\section{Introduction \label{sec1}}

The on going search for new physics (NP)
is mostly inspired by the
shortcomings of the SM in addressing some of the fundamental
questions in modern particle physics, such as the
hierarchy problem,
the flavor patterns in the fermionic sector and dark matter.
Some of these unresolved issues may be closely related
and may have TeV-scale origins, thus inspiring the search
for TeV-scale NP, both theoretically and experimentally.
Indeed, two seemingly unrelated interesting measurements of
both the ATLAS \cite{ATLAS1,ATLAS2} and the CMS \cite{CMS1,CMS2}
collaborations at CERN, have been recently reported:
\begin{enumerate}
\item A possible $(2-4) \sigma$ (local) excess in the diphoton invariant mass distribution
around 750 GeV,
corresponding to a signal cross-section roughly in the range
$\sigma(pp \to \gamma \gamma) \sim 3-13$ fb $(1\sigma)$, see e.g.,
\cite{1512.05777,1601.04751,1605.09401}.
The interpretation of this excess signal has a slight preference
to a spin 0 resonance, produced via gluon-fusion and having a total
width ranging from sub-GeV to 45 GeV, with a more significant signal
obtained in the ATLAS analysis for a scalar with a total width
$\Gamma \sim 45~{\rm GeV}$ \cite{ATLAS1}.
\item A possible $(1-2.5) \sigma$ excess in the measurement of the LFV decay
$h \to \tau \mu$ of the 125 GeV light Higgs. In particular,
the CMS collaboration finds $BR(h \to \tau \mu) = 0.84\%^{+0.39\%}_{-0.37\%}$ \cite{CMS2},
while the ATLAS collaboration finds $BR(h \to \tau \mu) = (0.53 \pm 0.51)\%$ \cite{ATLAS2}.
\end{enumerate}

Whether or not these two measurements are confirmed, it emphasizes the importance of the current
efforts in the search for NP, since it provides an interesting manifestation/example
of the exciting possibility that the building blocks of new TeV-scale physics may
have rather non-conventional properties, potentially with important repercussions for both flavor and the hierarchy
problems. For example, the new heavy scalar particle, $S$,
responsible for the 750 GeV $\gamma \gamma$ excess,
should have a rather narrow width and suppressed decay rates
into ``conventional" channels such as $S \to WW,~ZZ,~t  \bar t$,
for which no excess signals
has been observed within the currently available sensitivity of the corresponding LHC searches.
In addition, such a heavy scalar $S$ is most likely related
to the light 125 GeV scalar state and, therefore,
might also be involved in flavor changing (FC) transitions in the fermionic sector.
Such properties of the would be new 750 GeV resonating particle
are, therefore, very challenging to accommodate in models beyond the SM,
in particular, in supersymmetric models or in models that involve
extra space-time dimensions, which seem to have a more fundamental origin and, therefore,
likely linked to physics at higher energy scales.
Nonetheless, we will show in this paper that a certain class of low-energy effective 2HDM
frameworks with a 4th generation of heavy chiral fermions may be interesting
candidates for such ``exotic" TeV-scale NP.

In particular, since no evidence for such fundamentally structured theories has yet been seen,
a frequently adopted phenomenological approach in studies of NP,
is to construct TeV-scale models
which require a UV completion and may, thus, be
viewed as low energy effective frameworks for the underlying dynamics.
Such models are useful
as a guide for the exploration and model building of more fundamental theories and they
often include new heavy fermionic and scalar states with
sub-TeV masses. One of the simplest variants of an effective low-energy NP candidate, which dates back to the 1980's
\cite{SM4proc}, is the SM with an additional 4th generation of fermions; the so called SM4
(for useful reviews see e.g., \cite{sher0}).
Indeed, since three generations of chiral fermions have been observed in nature,
it is natural to ask why not four generations of chiral fermions?
It is quite interesting that this simple extension of the SM may address
some of the theoretical challenges in particle physics, such as:
electroweak symmetry breaking (EWSB) and the hierarchy problem \cite{DEWSB},
the CP-violation and the strength of the first-order phase transition needed
to explain the origin of matter - anti matter asymmetry in the universe \cite{baryo-ref,fok},
and flavor physics \cite{flavor}.
As is well known, the SM4 (i.e., with four generations of fermions and one Higgs doublet)
is now excluded, since it cannot accommodate
the measured SM-like properties of the 125
GeV scalar, see e.g., \cite{SM4-Higgs-bound,Lenz},
primarily due to an ${\cal O}(10)$ enhancement in the gluon-fusion light Higgs production mechanism
from diagrams with $t^\prime$ and $b^\prime$ in the loops \cite{Kribs_EWPT}; see, however, W.-S. Hou in \cite{1606.03732}.

This fact, along with the rather stringent direct limits on the masses
of such heavy quarks (to be discussed later), has led
to a common belief that generic extensions to the SM with
heavy chiral 4th generation fermions
$t^\prime,b^\prime,\nu^\prime,\tau^\prime$ are excluded.
However, as was suggested by us a few years ago
\cite{ourpaper1} and will be demonstrated again here,
this is not the case when the heavy
4th generation chiral sector is embedded in frameworks with an extended Higgs sector (see also \cite{1312.4249}).
Indeed, an extended Higgs sector in the context of 4th generation heavy fermions may come in handy
for further addressing flavor problems \cite{ourpaper1} and the strength of the EW phase transition required for
baryogenesis \cite{fok,1401.1827}.
In particular, we will consider in this paper a version of a 2HDM introduced by us in \cite{ourpaper1}
 - the so called 4G2HDM of type I, where
a chiral 4th generation doublet of heavy fermions (quark and lepton) is
added and is coupled {\it only} to one of the scalar doublets
(the ``heavy" doublet), while the SM 1st-3rd generations fermions
are coupled only to the other doublet (the ``light" doublet).
We will show in this paper that this 4G2HDM
is a well motivated and valid low-energy model,
which is compatible
with the 125 GeV signals (see also \cite{our125}), with PEWD and with the
existing direct bounds on the heavy fermions, and at the same time
can also accommodate the recent indications for a new 750 GeV scalar resonance
in the $\gamma \gamma$ channel.

As was shown in \cite{ourpaper1}, the price to pay when adding
another heavy SM-like chiral fermion doublet is that
such constructions posses a nearby threshold/cutoff at the several TeV scale,
which is manifest (as Landau poles) in the evolution of the Yukawa and Higgs
potential couplings \cite{ourpaper1,ourhybrid}.
Indeed, the large Yukawa couplings of the heavy chiral fermions can be thought of as a reflection
of an underlying TeV-scale strong dynamics, so that
the 4G2HDM framework should be viewed as a low energy (i.e., sub-TeV)
effective model of an underlying strongly interacting sector.
In particular, if the new heavy chiral fermions are viewed as the agents
of EWSB (and are, therefore, linked to
strong dynamics at the nearby TeV-scale,
see e.g., \cite{DEWSB-heavyF,luty}),
then more Higgs particles, which may be
composites of these 4th generation fermions, are expected at the sub-TeV
regime.$^{[1]}$\footnotetext[1]{Early
attempts in this direction investigated
the possibility of using the top-quark as the
agent of dynamical EWSB via top-condensation \cite{top-con}.
These models, however, fail to reproduce the observed value
of the top-quark mass. Moreover, as opposed to the case of condensates
of the heavy 4th generation fermions, where the typical cutoff for the new strong interactions
is of ${\cal O}(1)$ TeV,
the top-condensate models require
a corresponding cutoff many orders of magnitudes larger than $m_t$,
i.e., of ${\cal O}(10^{17})$ GeV, thus resulting in a severely fine-tuned
picture of dynamical EWSB.}
In such scenarios
the resulting low-energy effective
theory may contain more than a single composite Higgs field
\cite{ourpaper1,luty,DEWSB-multiH}
and may thus resemble a two (or more) Higgs doublet framework
(for other related studies of the phenomenology of multi-Higgs 4th generation models
see e.g., \cite{4G2HDM-others}).

The purpose of this work is to revisit
the 4G2HDM of \cite{ourpaper1}, studying its compatibility with the updated
measurements of the 125 GeV light Higgs signals and with
PEWD. We will also confront our model with
the 750 GeV $\gamma \gamma$ excess and study its compatibility with
a sub-percent branching ratio of the light Higgs in the FC decay channel $h \to \tau \mu$.
Indeed, many interesting and exotic constructions beyond the SM
have been suggested  as possible explanations
of the 750 GeV $\gamma \gamma$ excess (too many to be cited here); in most cases involving
new degrees of freedom beyond just the 750 GeV resonating particle.
In particular, the relevance of 2HDM frameworks to the 750 GeV $\gamma \gamma$
excess has been intensively studied in the past
several months, where it was shown
that the simplest 2HDM extension to the SM, in which
no additional heavy degrees of freedom are added
 (i.e., beyond the extended scalar sector), cannot
accommodate the necessary enhancement in $\sigma(pp \to H(750) \to \gamma \gamma)$,
see e.g., \cite{gilad}.
Consequently, extended 2HDM models with TEV scale vector-like (VL) fermions have been suggested
for addressing the 750 GeV resonance signal \cite{2HDMVLQ}.
The upshot of these studies is that, the needed
enhancement in the 1-loop production and decay channels $gg \to H(750)$
and $H(750) \to \gamma \gamma$, requires several
copies of VL fermions and/or VL fermions with charges appreciably larger
than those of the SM fermions, unless
their Yukawa couplings are much larger than one.
The 4G2HDM considered in this work
is, therefore, conceptually simpler, relying on new heavy
fermionic degrees of freedom with properties similar to the SM fermions
in a model that already exists in the literature.

The paper is organized as follows: in section 2 we describe the type I 4G2HDM and we layout
the physical parameters that are used in the numerical analysis. In section 3 we show our
results and in section 4 we discuss their phenomenological consequences. In section 5 we discuss our results and summarize.

\section{The 4G2HDM: a 2HDM with 4th generation fermions \label{sec2}}
Motivated by the idea that TeV-scale scalar degrees of freedom may emerge as composites associated with
heavy fermions, we assume that the low-energy (sub-TeV) effective framework is parameterized by a 2HDM
with a chiral SM-like 4th generation of heavy fermions. Specifically, the model is constructed following
\cite{ourpaper1}, such that
one of the Higgs fields ($\phi_h$ - the ``heavier" field)
couples only to the new heavy 4th generation fermionic fields,
while the second Higgs field ($\phi_\ell$ - the ``lighter" field)
is responsible for the mass generation of all other (lighter) fermions (i.e., the 1st-3rd generation SM fermions).
In this model, named in \cite{ourpaper1} the 4G2HDM of type I (here we will refer to it simply as the 4G2HDM),
the Yukawa interaction Lagrangian can be realized in terms of a $Z_2$-symmetry
under which the fields transform as follows:
\begin{eqnarray}
  \Phi_{\ell}\to-\Phi_{\ell},~ \Phi_{h}\to+\Phi_{h},~ F_{L}\to+F_{L}~,
  f_{R}\to-f_{R}\;(f={\rm SM~fermions}),~f^\prime_{R}\to +f^\prime_{R}\;(f^\prime={\rm 4th~gen.~fermions})~,
\label{eq:z2}
\end{eqnarray}
where $F_L$ and $f_{R},f_R^\prime$ are the SU(2) fermion (quark or lepton) doublets and singlets, respectively,
and $\Phi_{\ell,h}$ are the two Higgs doublets
$\Phi_i =\left( \phi^{+}_i,\frac{v_i+\phi^{0}_i}{\sqrt{2}} \right)$, $i=\ell,h$.

The Yukawa potential that respects the above $Z_2$-symmetry  is:
%
\begin{eqnarray}
\mathcal{L}_{Y}= -\bar{F}_{L}
\left( \Phi_{\ell} Y_d^f \cdot \left( I-{\cal I}
\right) +
\Phi_{h}Y_d^f \cdot {\cal I} \right) f_{d,R}
-\bar{F}_{L}
\left( \tilde\Phi_{\ell} Y_u^f \cdot \left( I - {\cal I} \right) +
\Phi_{h} Y_u^f \cdot {\cal I} \right)
f_{u,R} + h.c.\mbox{ ,}
\label{eq:LY4G}
\end{eqnarray}
%
where $f_{u,R}$ and $f_{d,R}$ are the up and down-type SU(2) fermion singlets (quark or lepton of all four generations),
$I$ is the identity matrix and ${\cal I}$ is the diagonal $4\times4$ matrix
${\cal I} \equiv {\rm diag}\left(0,0,0,1\right)$.

The scalar sector contains five massive states: a charged scalar $H^+$, a CP-odd
state $A$ and two CP-even scalars $h,H$, so that $h$ is the lighter one,
corresponding to the observed 125 GeV Higgs boson. These physical states are
related to the components of the two SU(2) scalar doublets via:
\begin{eqnarray}
H= s_\alpha {\rm Re} \left( \phi_h^0 \right) + c_\alpha {\rm Re}\left( \phi_\ell^0 \right)
~&,&~
A= s_\beta {\rm Im} \left( \phi_\ell^0 \right) - c_\beta {\rm Im}\left( \phi_h^0 \right)
~, \nonumber \\
h= c_\alpha {\rm Re}\left(\phi_h^0\right) - s_\alpha {\rm Re}\left(\phi_\ell^0\right)
~&,&~
H^+= s_\beta \phi_\ell^+  - c_\beta \phi_h^+
~, \label{Higgsangles}
\end{eqnarray}
where $s_\alpha(c_\alpha)=\sin\alpha(\cos\alpha)$, $\alpha$ being the Higgs mixing angle in the CP-even sector and $s_\beta(c_\beta)=\sin\beta(\cos\beta)$, where $\tan\beta \equiv v_h/v_\ell$ is the
ratio between the VEV's of the heavy and light Higgs fields.

The Yukawa Higgs-quark-quark interactions in the 4G2HDM are (similar
terms can be written for the leptons) \cite{ourpaper1}:
\begin{eqnarray}
{\cal L}(h q_i q_j) &=& \frac{g}{m_W \sin 2\beta} \bar q_i \left\{ m_{q_i} s_\alpha s_\beta \delta_{ij}
- \cos(\beta -\alpha) \cdot
\left[ m_{q_i} \Sigma_{ij}^q R + m_{q_j} \Sigma_{ji}^{q \star} L \right] \right\} q_j h \label{Sff1}~, \\
{\cal L}(H q_i q_j) &=& \frac{g}{m_W \sin 2\beta} \bar q_i \left\{ -m_{q_i} c_\alpha s_\beta \delta_{ij}
+ \sin(\beta -\alpha)\cdot
\left[ m_{q_i} \Sigma_{ij}^q R + m_{q_j} \Sigma_{ji}^{q \star} L \right] \right\} q_j H ~, \\
{\cal L}(A q_i q_j) &=& - i I_q \frac{g}{m_W \sin 2\beta} \bar q_i \left\{ m_{q_i} s_\beta^2 \gamma_5 \delta_{ij}
- \left[ m_{q_i} \Sigma_{ij}^q R - m_{q_j} \Sigma_{ji}^{q \star} L \right] \right\} q_j A ~,
\end{eqnarray}
\begin{eqnarray}
{\cal L}(H^+ u_i d_j) = \sqrt{2} \frac{g}{ m_W \sin 2\beta} \bar u_i \left\{
\left[ m_{d_j} s_\beta^2 \cdot V_{u_id_j} - m_{d_k}
V_{ik} \Sigma^{d}_{kj} \right] R + \left[ -m_{u_i} s_\beta^2 \cdot V_{u_id_j} + m_{u_k}
\Sigma^{u \star}_{ki} V_{kj} \right] L
 \right\} d_j H^+ \label{Sff2}~,
\end{eqnarray}
where $V$ is the $4 \times 4$ CKM matrix, $q=d$ or $u$ for down or up-quarks with $I_d=-1$ and $I_u=+1$, respectively,
and $R(L)=\frac{1}{2}\left(1+(-)\gamma_5\right)$.
Also, $\Sigma^d$ and $\Sigma^u$ are new mixing matrices where
all FCNC effects of the 4G2HDM are encoded. They are
obtained after diagonalizing the quark mass matrices and, therefore, depend on
the rotation (unitary) matrices of the right-handed down and up-quarks
$D_R$ and $U_R$, respectively. In particular, for ${\cal I} \equiv {\rm diag}\left(0,0,0,1\right)$
in Eq.~\ref{eq:LY4G}, we have (see \cite{ourpaper1}):$^{[2]}$\footnotetext[2]{Note that this
is in contrast to
``standard" frameworks such as the SM and the 2HDM's of types I and II,
where the right-handed mixing matrices $U_R$ and $D_R$ are non-physical,
being ``rotated away" in the diagonalization procedure of the quark masses.}
\begin{eqnarray}
\Sigma_{ij}^d = D_{R,4i}^\star D_{R,4j} ~,~ \Sigma_{ij}^u  =U_{R,4i}^\star U_{R,4j} ~.
\label{sigma}
\end{eqnarray}

The Yukawa structure and couplings defined by Eqs.~\ref{eq:LY4G}-\ref{sigma} is assumed
to be copied to the leptonic sector, see \cite{ourg-2}.
In the following sections \ref{sec3} and \ref{sec4},
for illustrative purposes (and without loss of generality), we will set
$\Sigma^{d,u} \to {\rm diag}\left(0,0,0,1\right)$ in both the quark and lepton sectors,
so that FCNC effects (in particular, between the 4th generation fermions
and the SM fermions) are ``turned off".
In fact, from the phenomenological point of view, it is sufficient
to assume that $\Sigma^{u}_{34,43} \to 0$ (i.e., forbidding
the decay $t^\prime \to t h$) and $V_{i4,4i} \to 0$ ($i=1,2,3$, thus forbidding
the decays $t^\prime \to d_i W$ and $b^\prime \to u_i W$ with $d_i=d,s,b$ and $u_i=u,c,t$)
in order to accommodate relatively light $t^\prime$ and $b^\prime$ with masses as low as 350 GeV,
since the existing stringent exclusion limits of $m_{t^\prime},m_{b^\prime} \gsim 700$ GeV,
are based on searches that assume 100\% branching ratios of the 4th generation
quarks into one of the channels: $t^\prime \to th,tZ,d_i W$ and
$b^\prime \to Zb,u_iW$ \cite{pdg,1509.04261}.
We will, therefore, assume that
the dominant $t^\prime$ and $b^\prime$ decays are into one of the FC channels
$t^\prime \to u_ih$ and
$b^\prime \to d_i h$ ($u_i=u,c$ and $d_i=d,s,b$), due to small
FCNC entries in $\Sigma^{u,d}$ (which have no effect on the results presented in sections \ref{sec3} and \ref{sec4}),
in which case small off-diagonal CKM entries $V_{14,41}$ and/or $V_{24,42}$ are also allowed
as long as $BR(t^\prime \to d_i W),~BR(b^\prime \to u_iW) \lsim 0.5$ \cite{1509.04261}.
Such flavor structures, may have interesting phenomenological implications,
as will be discussed in section \ref{sec5}.

The 2HDM scalar sector is parameterized by seven free parameters (after minimization
of the potential), which, in the so called
``physical basis", can be chosen as the four physical Higgs masses ($m_h,~m_H,~m_A,~m_{H^+}$),
the two angles
$\beta$ and $\alpha$ and one parameter from the scalar potential, which is needed
in order to specify the scalar couplings, in particular,
$hH^+H^-$ (which enters in the 1-loop $h \to \gamma \gamma$ decay), $HH^+H^-$
(which enters the 1-loop $H \to \gamma \gamma$ decay)
and $Hhh$ (required for the decay $H \to hh$).
In the physical basis, these
scalar couplings can be written at tree-level as
(see e.g., \cite{0408364}):
\begin{eqnarray}
\lambda_{Hhh}=-\frac{\cos(\alpha-\beta)}{2 v \sin 2\beta}
\left[ \sin 2 \alpha \left(m_h^2+2m_H^2 \right) -
\left( 3 \sin 2 \alpha - \sin 2 \beta \right)
\frac{m_{\ell h}^2}{s_\beta c_\beta} \right] \label{lam1} ~,
\end{eqnarray}
\begin{eqnarray}
\lambda_{hH^+H^-}=-\frac{1}{2 v \sin 2\beta}
\left[ \left( \cos(\alpha-3\beta)+3 \cos(\alpha+\beta) \right) m_h^2 -
4 \sin 2 \beta \sin(\alpha-\beta) m_{H^\pm}^2 -
4 \cos(\alpha+\beta) \frac{m_{\ell h}^2}{s_\beta c_\beta} \right] \label{lam2}  ~,
\end{eqnarray}
\begin{eqnarray}
\lambda_{HH^+H^-}=-\frac{1}{2 v \sin 2\beta}
\left[ \left( \sin(\alpha-3\beta)+3 \sin(\alpha+\beta) \right) m_H^2 +
4 \sin 2 \beta \cos(\alpha-\beta) m_{H^\pm}^2 -
4 \sin(\alpha+\beta) \frac{m_{\ell h}^2}{s_\beta c_\beta} \right] \label{lam3}  ~,
\end{eqnarray}
where $m_{\ell h}^2$ is a mass-like term,
$m_{\ell h}^2 \Phi_\ell^\dagger \Phi_h + h.c.$,
which softly breaks the above
$Z_2$-symmetry (i.e., $\Phi_{\ell}\to-\Phi_{\ell},~ \Phi_{h}\to+\Phi_{h}$),
and which can be used to specify the above tree-level
scalar couplings.

However, since
the working assumption of the 4G2HDM is that
the scalar sector may be strongly interacting at the near by few TeV scale,
the scalar potential is expected to be subject to significant renormalization and threshold effects.
Thus, the above scalar couplings are expected
to deviate from their tree-level values, depending
on the details of the UV completion and on the masses
of the heavy degrees of freedom of this model, see e.g.,
\cite{0408364,loop-cor}.
As an example, consider the 1-loop corrections to the $Hhh$ coupling $\lambda_{Hhh}$,
for $|\alpha| \to \pi/2$, in which case there is
no mixing between the light and heavy Higgs fields (see Eq.~\ref{Higgsangles}),
as
required in order to accommodate
the 750 GeV $\gamma \gamma$ excess in the 4G2HDM (see section \ref{sec4}).
In this limit, the Yukawa couplings of the 4th generation fermions
to the light Higgs state $h$ (i.e., $t^\prime t^\prime h$) vanish (see Eq.~\ref{Sff1} and Table \ref{tab4}) and
we find that
the dominant effect arises from the 1-loop triangle diagram
with the charged Higgs exchange in the loop, giving
a ``renormalized" $Hhh$ coupling $\bar\lambda_{Hhh} \equiv a_{Hhh} \lambda_{Hhh}$,
with:
\begin{eqnarray}
a_{Hhh} \approx 1+ \frac{m_{\ell h}^4}{m_H^2 v^2}
\frac{\left( 1-2c_\beta^2 \frac{m_{H^+}^2}{m_{\ell h}^2} \right)
\left(1+c_\beta^2 \frac{m_{H}^2}{m_{\ell h}^2} - 2s_\beta^2 \frac{m_{H^+}^2}{m_{\ell h}^2}
\right)}{2 \pi^2 (\sin2\beta)^2} I\left(m_h,m_H,m_{H^+} \right)~,
\end{eqnarray}
where $I\left(m_h,m_H,m_{H^+} \right)$ is the charged Higgs triangle loop integral, given by:
\begin{eqnarray}
I\left(m_h,m_H,m_{H^+} \right) = - \int_0^1 dx \int_0^{1-x} dy ~
\frac{1}{ (x+y) (x+y-1) m_h^2 - xy m_H^2 + m_{H^+}^2  } ~.
\end{eqnarray}
In particular, one roughly finds $|a_{Hhh}| \in \left\{0, 2\right\}$ when
$m_{H^+} \in \left\{500~{\rm GeV}, 1~{\rm TeV} \right\}$ and with
$m_H =750$ GeV, $m_h =125$ GeV and $m_{\ell h} \sim {\cal O}(1~{\rm TeV})$.
For example, $a_{Hhh} \sim -0.15$ for $m_{H^+} = m_H = 750$ GeV and
$m_{\ell h} =1.2$ TeV. In what follows we will, therefore,
define the ``renormalized" scalar couplings as: $\bar\lambda_i \equiv a_i \lambda_i$,
where $\lambda_i$ ($i=Hhh,~ hH^+H^-,~HH^+H^-$) are
the corresponding tree-level couplings in Eqs.~\ref{lam1}-\ref{lam3}, and
$a_i$ will be treated as free-parameters in the fit that will be varied in the range
$|a_i| \in \left\{0, 2\right\}$.

\section{The 125 GeV Higgs signals and PEWD \label{sec3}}

The measured signals of the 125 GeV Higgs particle, which in the 4G2HDM
is the light Higgs $h$, and PEWD impose stringent constraints
on the free parameter space of the 4G2HDM.
For the 125 GeV Higgs signals we use the measured values of the
``signal strength" parameters, which are defined as the ratio between the measured
rates and their SM expectation.
In particular, for a specific production and decay
channel $i \to h \to f$, the signal strength is defined as:
\begin{eqnarray}
\mu_{i}^f \equiv \mu_i \cdot \mu^f ~,
\end{eqnarray}
with
\begin{eqnarray}
\mu_i = \frac{\sigma(i \to h)}{\sigma(i \to h)_{SM}} = k_i^2~,~
\mu^f = \frac{BR(h \to f)}{BR(h \to f)_{SM}} = \frac{k_f^2}{R^T} ~,
\end{eqnarray}
where $k_j$ is the 4G2HDM coupling involved in $j \to h$ or $h \to j$ production or decay processes, normalized by its SM value, and $R^T$
is the ratio between the total width of $h$ in the 4G2HDM and the
total width of the SM 125 GeV Higgs. In particular,
\begin{eqnarray}
k_j \equiv \frac{k_j^{4G2HDM}}{k_j^{SM}} ~,~
R^T \equiv \frac{\Gamma_{h_{4G2HDM}}^{Total}}{\Gamma_{h_{SM}}^{Total}}~,
\end{eqnarray}
so that $\mu_i^f = k_i^2 k_f^2 / R^T$.

In Table \ref{tab1} we list the latest combined ATLAS and CMS six parameter fit from RUN1
\cite{ATLAS-CMS-125res}, of the measured
values for $\mu_{gg}^{\gamma \gamma},~\mu_{gg}^{WW^\star},
~\mu_{gg}^{ZZ^\star},~\mu_{gg}^{bb},~\mu_{gg}^{\tau \tau}$
and $\mu_V/\mu_{gg}$, where $\mu_V$ stands for Higgs
production via vector-boson fusion (VBF) or in association with a
vector-boson (VH).$^{[3]}$\footnotetext[3]{We neglect Higgs production
via $pp \to tth$ which, although included in the fit, is
2-3 orders of magnitudes smaller than the gluon-fusion channel}.
We also write in Table \ref{tab1} the model predictions for the various signal strengths in terms
of the normalized couplings defined above.

\begin{table}[htb]
\begin{center}
\begin{tabular}{c||c|c|}
 & measured value & model prediction / couplngs \\
\hline \hline
 $\mu_{gg}^{\gamma \gamma} $ & $1.13^{+0.24}_{- 0.21}$  & $k_g^2  k_\gamma^2 /R^T $ \\
\hline
$\mu_{gg}^{ZZ^\star}  $ & $1.29^{+0.29}_{- 0.25}$ & $k_g^2 k_V^2 /R^T$ \\
\hline
$\mu_{gg}^{WW^\star} $ & $1.08^{+0.22}_{- 0.19}$ & $k_g^2  k_V^2 /R^T$ \\
\hline
$\mu_{gg}^{bb} $ & $0.65^{+0.37}_{- 0.28}$ & $k_g^2 k_b^2/R^T$ \\
\hline
$\mu_{gg}^{\tau \tau} $ & $1.07^{+0.35}_{- 0.28}$ & $k_g^2  k_\tau^2 /R^T$ \\
\hline
$\mu_V/\mu_{gg} = $ & $1.06^{+0.35}_{- 0.27}$ & $k_V^2 / k_g^2$ \\
\hline
\end{tabular}
\caption{Measured
values \cite{ATLAS-CMS-125res} and model predictions in terms of normalized couplings (see text)
of the various production and decay channels for the 125 GeV Higgs,
using the signal strength prescription.
Note that while $k_V,k_b$ and $k_\tau$ are ratios of tree-level
couplings, $k_g$ and $k_\gamma$ are the normalized (with respect to the SM)
1-loop 4G2HDM couplings $hgg$ and $h\gamma \gamma$, respectively, calculated
using the formula in \cite{HHG}.
Also, in our 4G2HDM $k_W=k_Z=k_V$.}
\label{tab1}
\end{center}
\end{table}

For the PEWD constraints on the 4G2HDM, we update our study in \cite{ourpaper1}.
In particular, the effects of any new physics can be divided into those
which do and which do not couple directly
to the ordinary SM fermions.
For the former, the leading effect in the 4G2HDM comes from the decay
$Z \to b \bar b$, which is mainly sensitive to the
$H^+ t^\prime b$ and $W^+ t^\prime b$ couplings through one-loop exchanges of $H^+$ and $W^+$, as was analyzed in detail in \cite{ourpaper1}.
These contributions to $Z \to b \bar b$ are, however, absent in the currently studied
versions of the 4G2HDM, since our working assumption here
is that $V_{t^\prime b} \to 0$ and $\Sigma^{d,u} \to {\rm diag}\left(0,0,0,1\right)$,
so that the $H^+ t^\prime b$ and $W^+ t^\prime b$ vertices vanish
or are negligibly small (see previous section).

The effects which do not involve direct couplings to the ordinary fermions,
can be analyzed in the formalism of the oblique parameters S,T and U
\cite{peskin}. The contribution of a 2HDM with a 4th generation of chiral fermions
to the oblique parameters were studied in \cite{ourpaper1}.
This includes the pure 1-loop Higgs exchanges
to the gauge-bosons 2-point functions and the 1-loop exchanges of
$t^\prime$ and $b^\prime$ which shift the T parameter
and which involve the new SM4-like diagonal coupling
$W t^\prime b^\prime$ (here also the contributions involving the
off-diagonal couplings $W t^\prime b$ and $W t b^\prime$ are absent
since we assume $V_{t^\prime b},~ V_{t b^\prime} \to 0$, see also \cite{deltaT}).
These are calculated with respect
to the SM values and are bounded by a global fit to PEWD \cite{gfitter}:
\begin{eqnarray}
\Delta S &=& S - S_{SM} = 0.06 \pm 0.09 ~, \nonumber \\
\Delta T &=& T - T_{SM} = 0.1 \pm 0.07 \label{SandT}~,
\end{eqnarray}
with a correlation coefficient of $\rho = +0.91$.
These values are obtained for $\Delta U=0$
(the $U$ parameter is often set to zero since
it can be neglected in most new physics models and, in particular
in our 4G2HDM) and
with the SM reference values
$M_{H,{\rm ref}}=125$ GeV and $m_{t,{\rm ref}}=173$ GeV.
We, thus,
consider below the constraints from the 2-dimensional ellipse in the $S-T$ plane
which, for a given confidence level (CL), is defined by:
\begin{widetext}
\begin{eqnarray}
\left(
\begin{array}{c} S - S_{exp} \\ T - T_{exp}  \end{array}
\right)^T
\left(
\begin{array}{cc} \sigma_S^2 & \sigma_S \sigma_T \rho \\
\sigma_S \sigma_T \rho & \sigma_T^2 \end{array}
\right)
\left(
\begin{array}{c} S - S_{exp} \\ T - T_{exp}  \end{array}
\right)
= - 2 {\rm ln}\left( 1 - CL \right) \label{ST2}~,
\end{eqnarray}
\end{widetext}
where $S_{exp} = 0.06$ and $T_{exp} = 0.1$ are the best fitted (central)
values, $\sigma_S = 0.09, \sigma_T = 0.07$
are the corresponding standard deviations and $\rho=0.91$
is the (strong) correlation factor between S and T.

We thus perform a random (``blind") scan of the
relevant parameter space, imposing compatibility at 95\% CL
of the 4G2HDM with the measured 125 GeV Higgs signals
listed above and with the best fitted values of $S$ and $T$
using Eqs.~\ref{SandT} and \ref{ST2}. In particular, we fix $m_H=750$ GeV
(for compatibility with the recent 750 GeV $\gamma \gamma$ signal, see next section) and scan
the rest of the parameters over the following ranges:
\begin{eqnarray}
\alpha \in \left[ - \frac{\pi}{2},\frac{\pi}{2} \right] ~,~
\tan\beta \in \left[ 0.4 , 10 \right]~,~ a_i\in \left[ -2,2 \right] ~
(i=hH^+H^-,HH^+H^-,Hhh) ~, \nonumber
\end{eqnarray}
\begin{eqnarray}
m_{\ell h}^2 \in \left[ - \left(2~ {\rm TeV} \right)^2,\left(2~ {\rm TeV} \right)^2 \right] ~,~
m_{A,H^+} \in \left[ 300~ {\rm GeV} ,1.5~ {\rm TeV} \right]  ~, \nonumber
\end{eqnarray}
\begin{eqnarray}
m_{t^\prime,b^\prime} \in \left[ 350~{\rm GeV} , 500~{\rm GeV} \right] ~,~
m_{\nu^\prime,\tau^\prime} \in \left[ 200~{\rm GeV} , 1200~{\rm GeV} \right] ~.
\end{eqnarray}

\begin{figure}
\begin{center}
\includegraphics[scale=0.27]{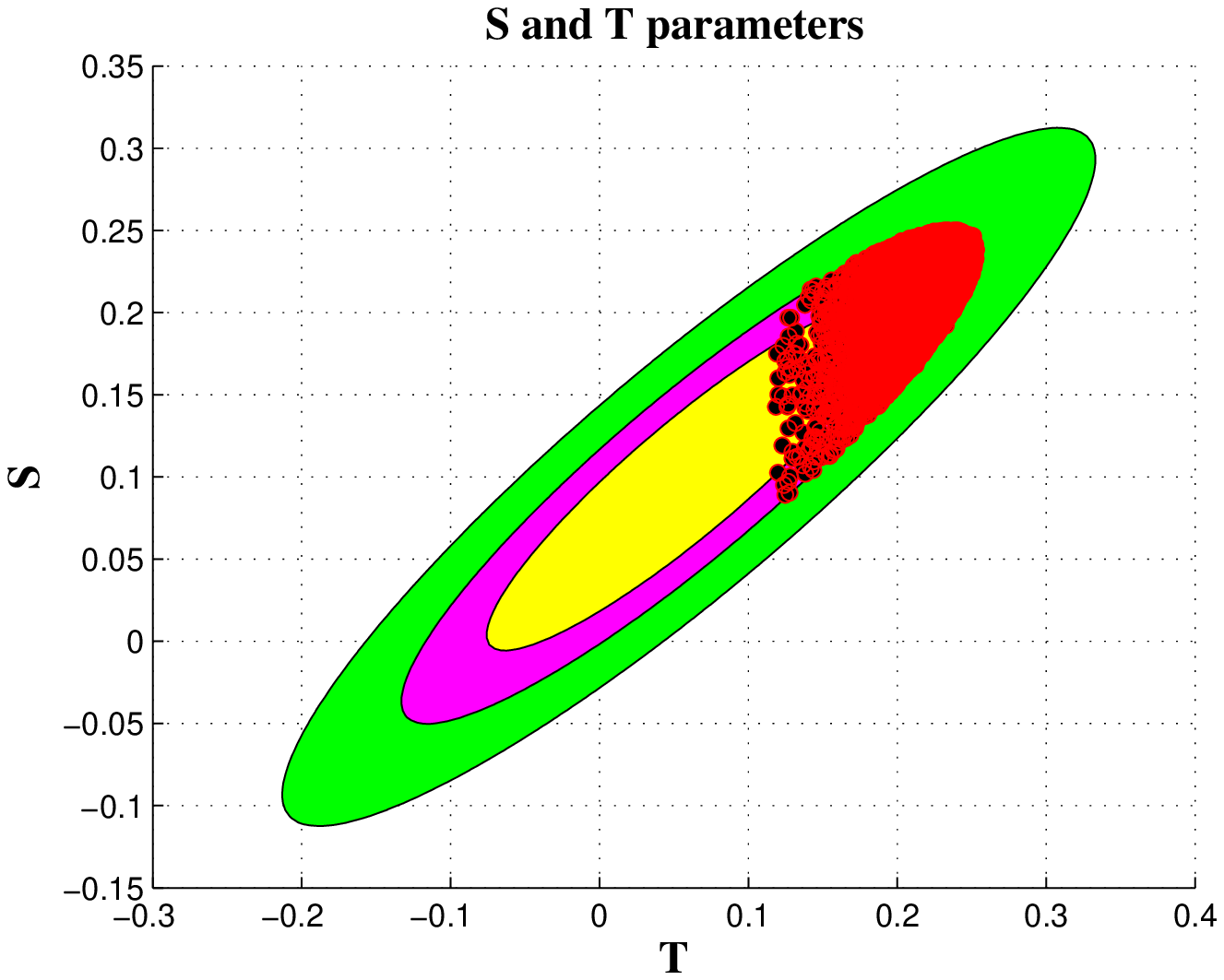}
\includegraphics[scale=0.27]{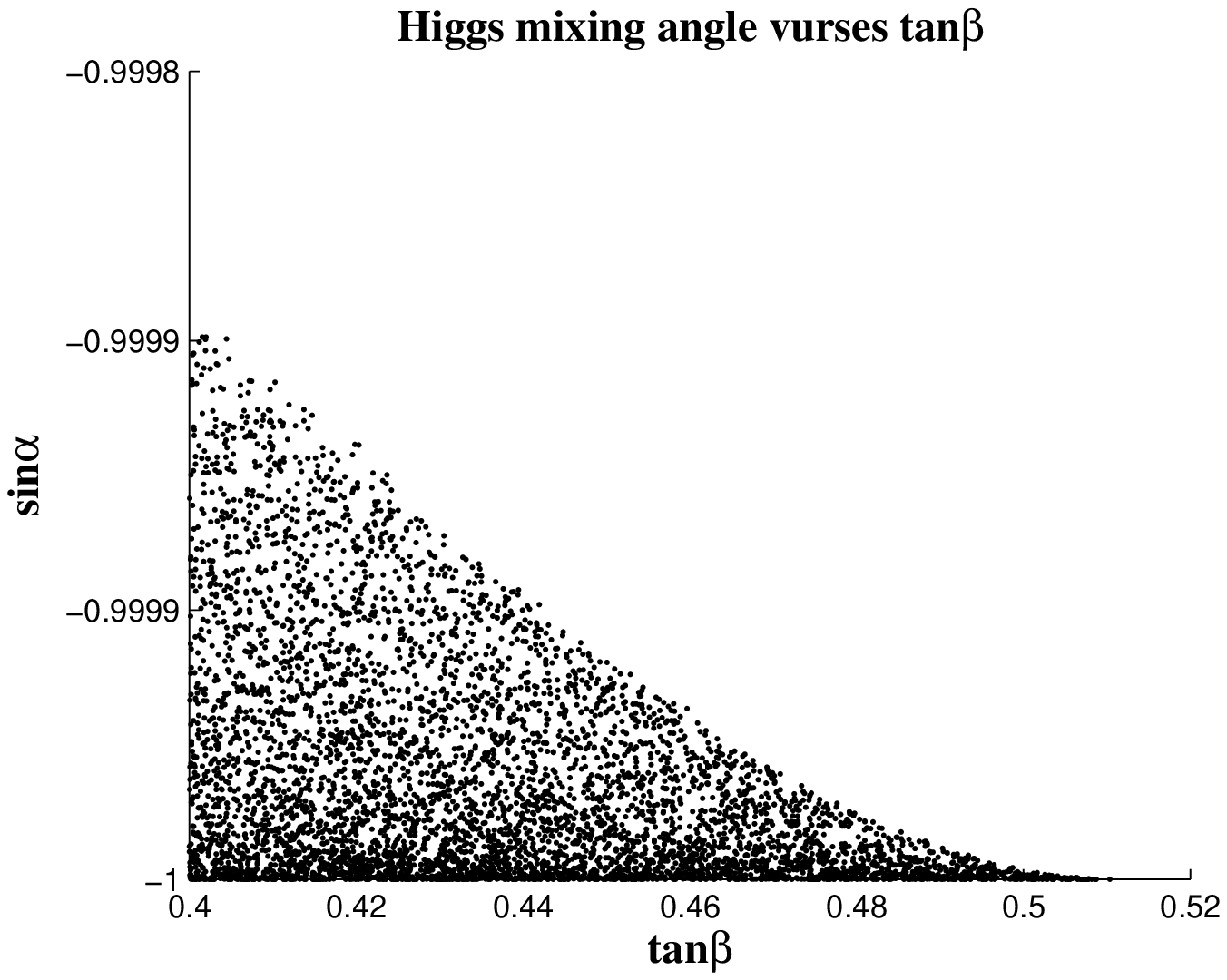}
\includegraphics[scale=0.27]{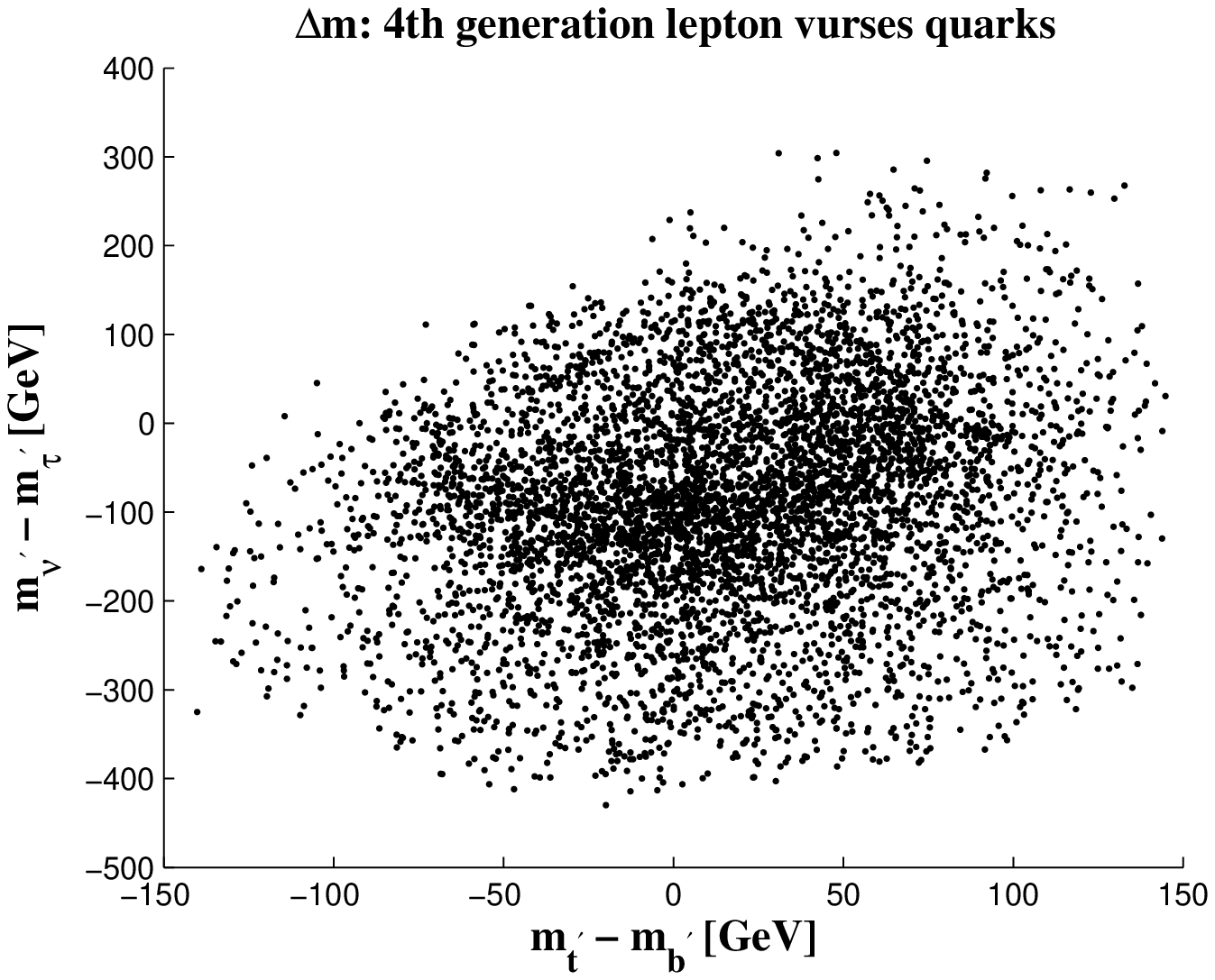}
\includegraphics[scale=0.27]{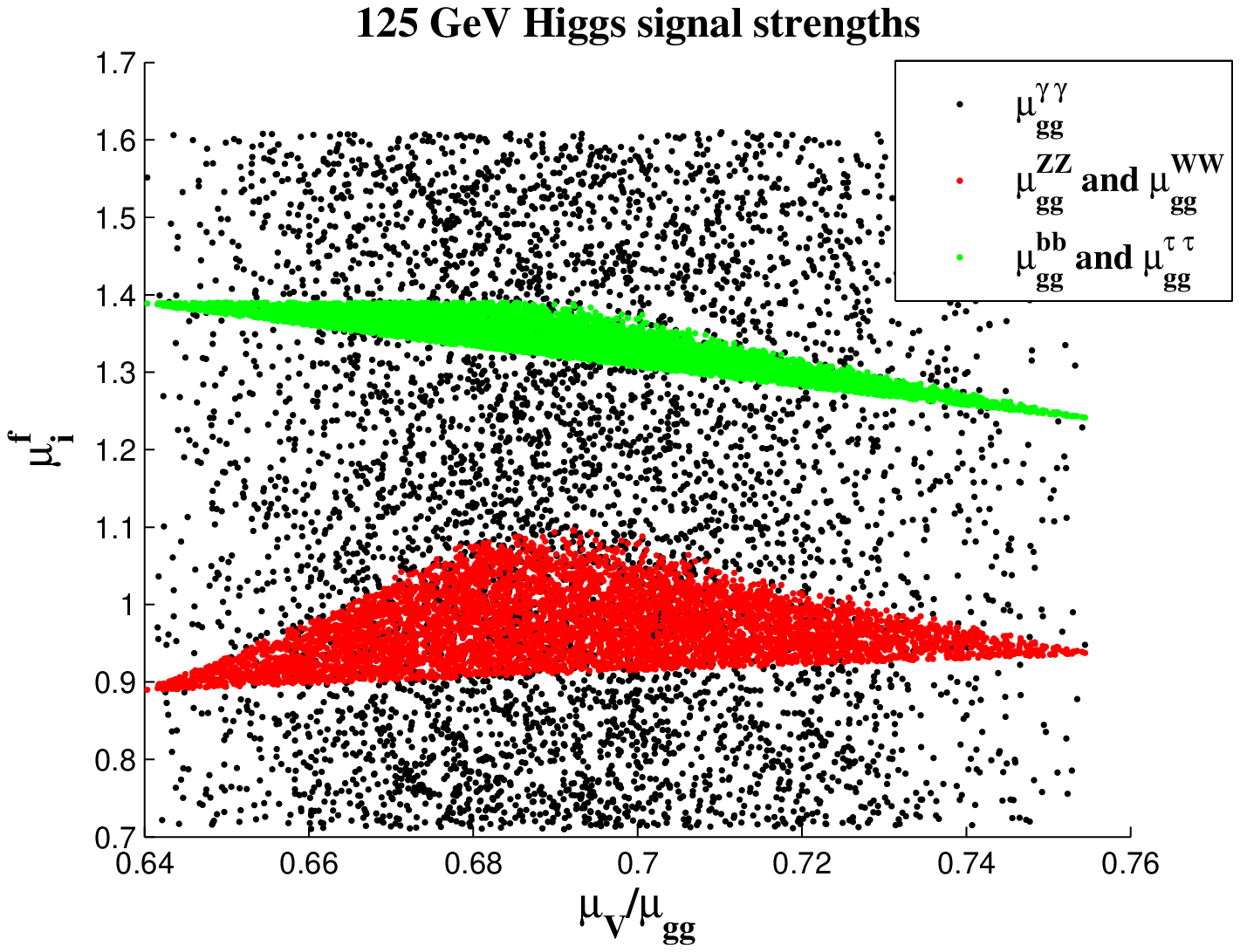}
\includegraphics[scale=0.27]{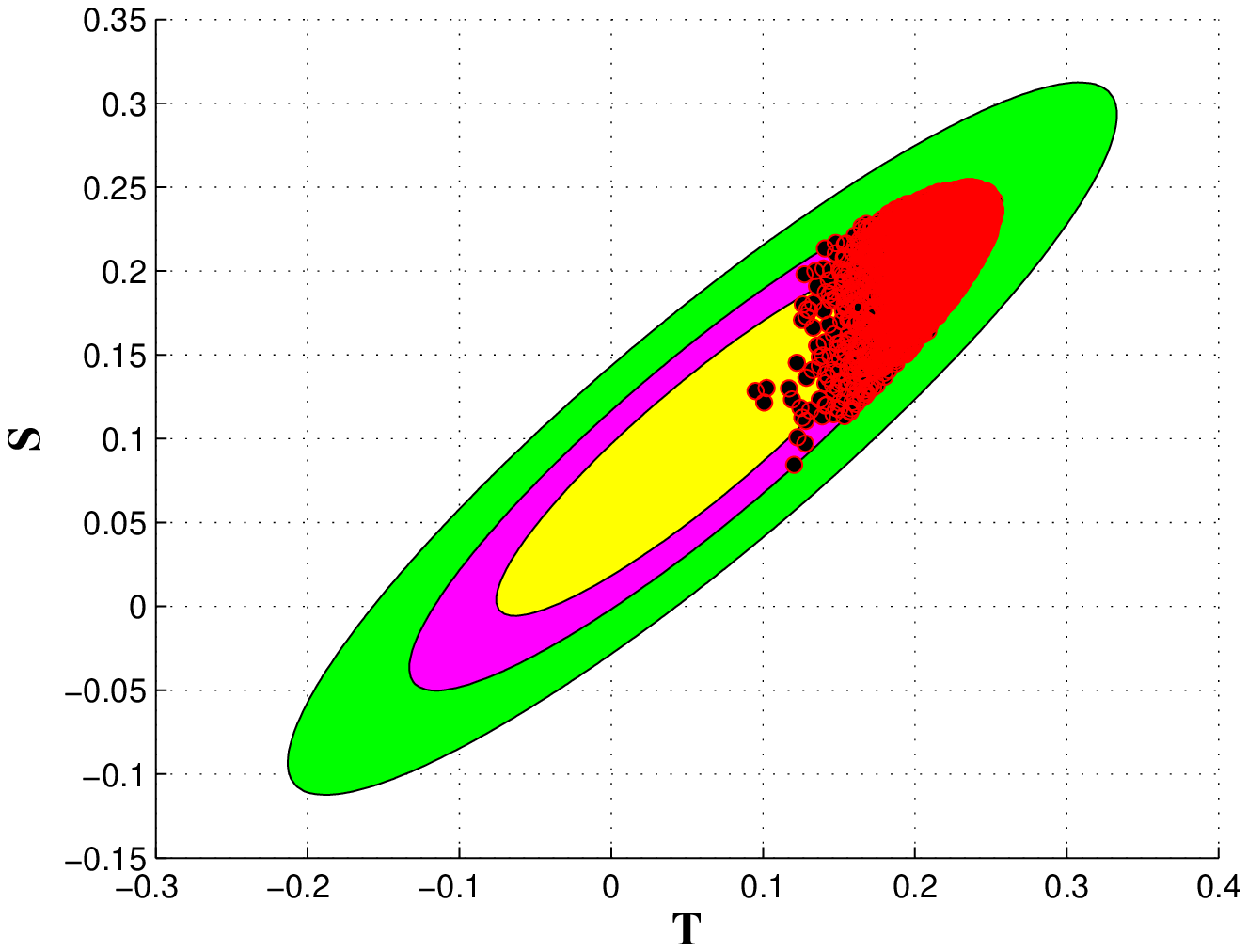}
\includegraphics[scale=0.27]{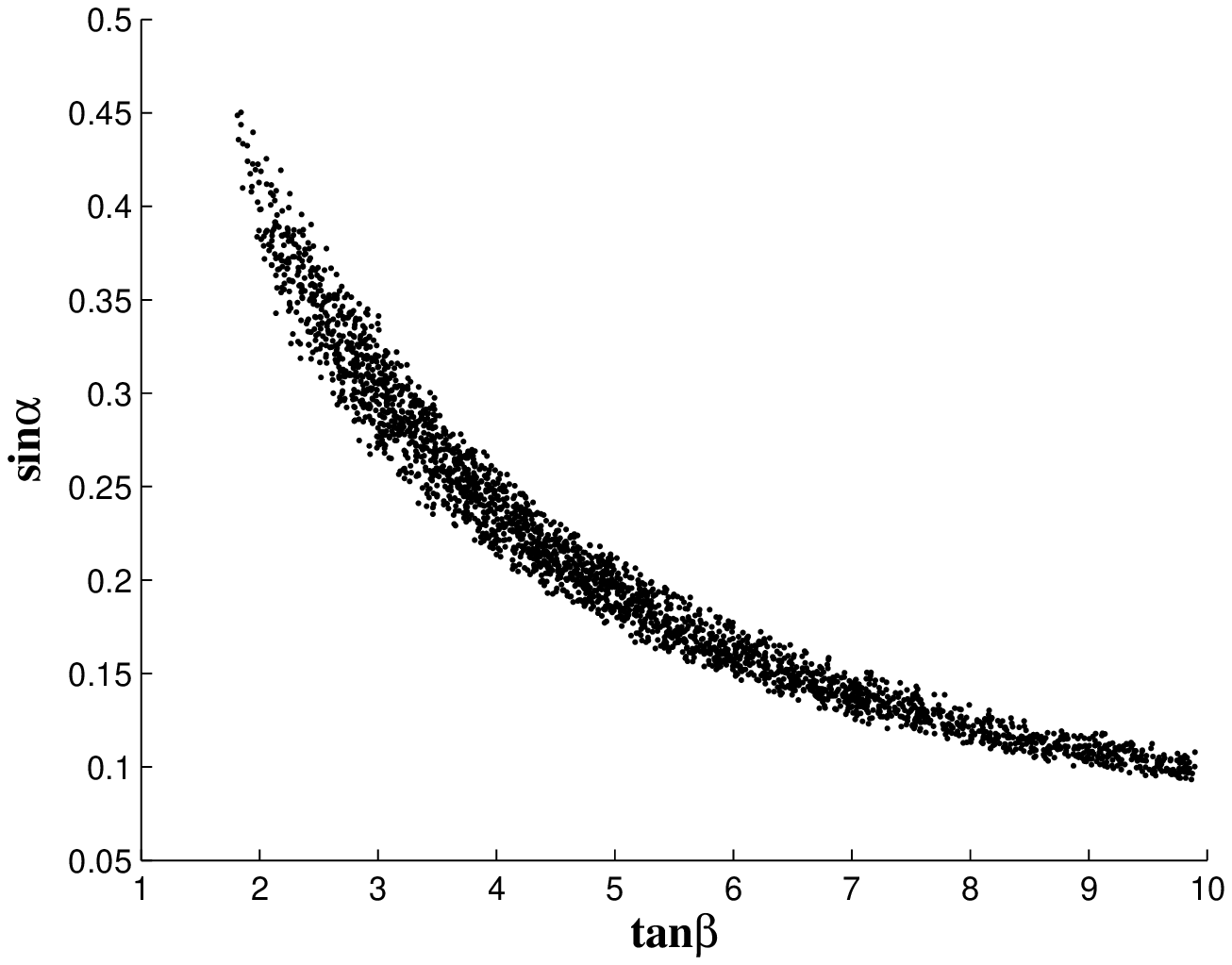}
\includegraphics[scale=0.27]{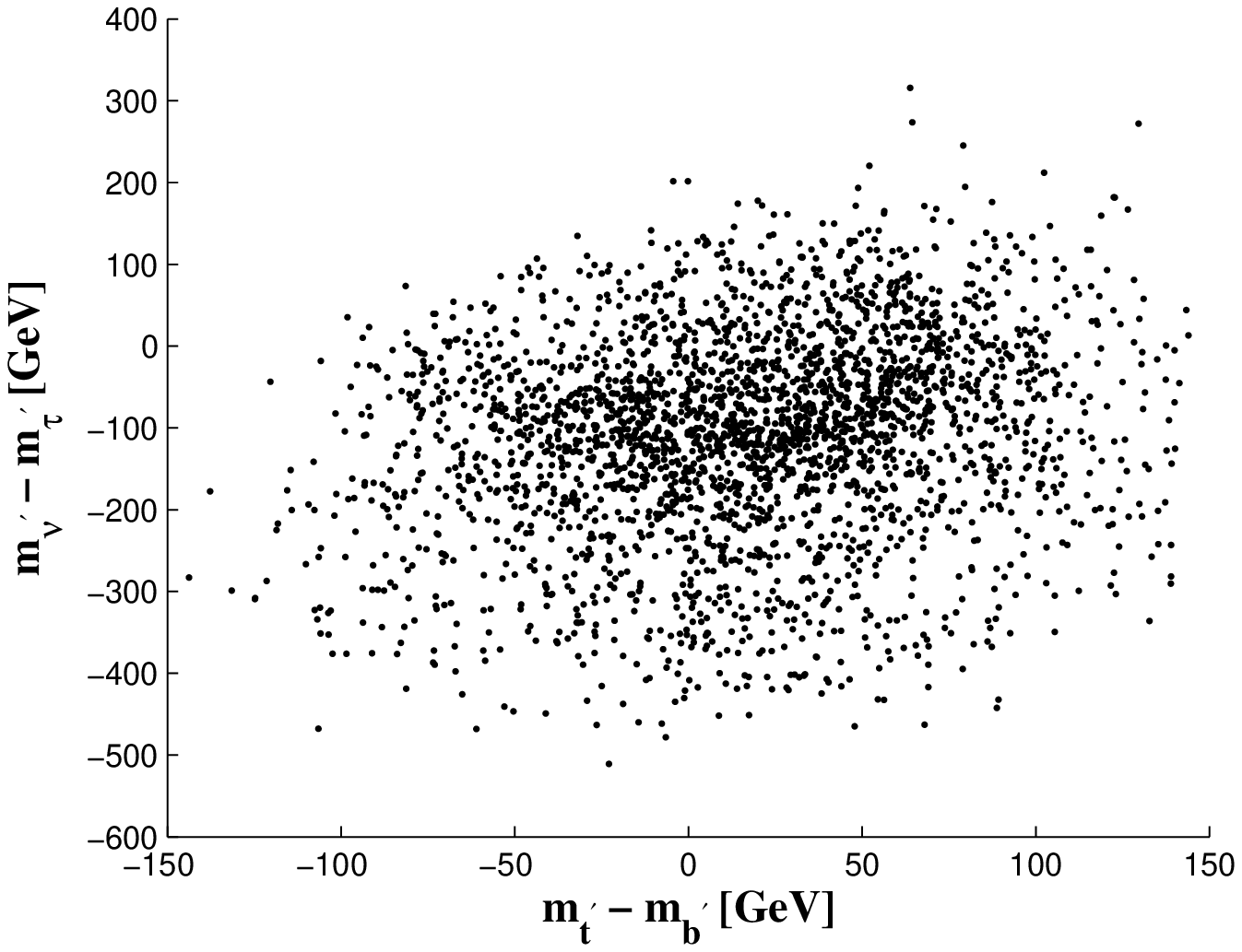}
\includegraphics[scale=0.27]{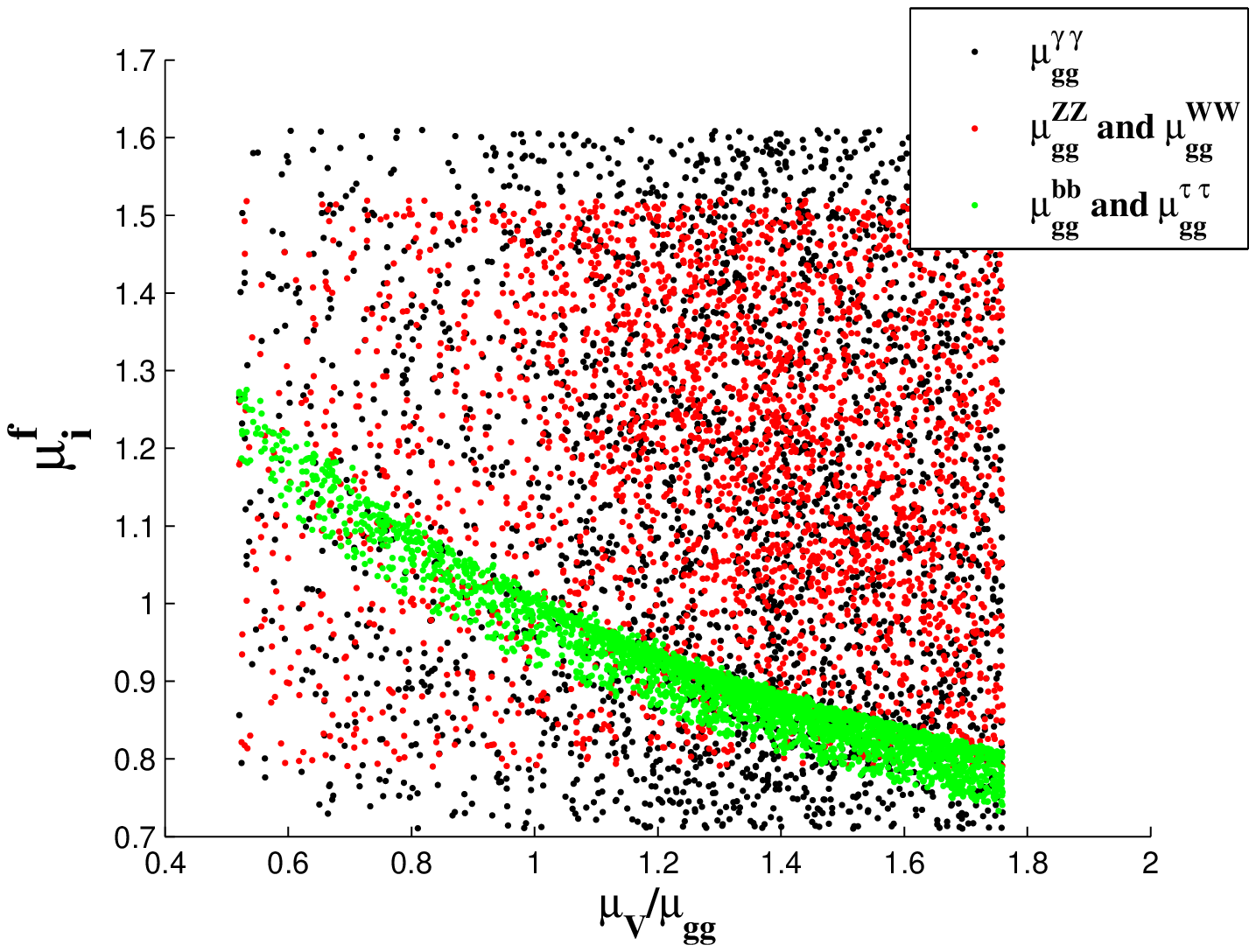}
\end{center}
\caption{The distribution of the 4G2HDM parameter space that is compatible
with the 125 GeV signals and PEWD. From left to right: in the $S -T$ plane (yellow, pink and green ellipses correspond
to the 68\%, 95\% and 99\% CL allowed contours, respectively), in
the $\tan\beta - \sin\alpha$ plane, in the
$\Delta m_{\ell^\prime} - \Delta m_{q^\prime}$ plane,
where $\Delta m_{\ell^\prime} \equiv m_{\nu^\prime} - m_{\tau^\prime}$ and $\Delta m_{q^\prime} \equiv m_{t^\prime} - m_{b^\prime}$
and the corresponding 125 GeV signal
strengths (on right). Case 1 in the upper row, case 2 in the middle row and case 3 in the lower row,
see text.
\label{fig1}}
\end{figure}

We find two types of possible 4G2HDM ``solutions":
\begin{description}
\item{case 1:} $\tan\beta \leq 0.5$, $\sin\alpha \to -1$ and $m_{\ell h}^2 > 0$.
\item{case 2:} $\tan\beta \geq 2$, $\sin\alpha \sim 0.1 - 0.45$ and any $m_{\ell h}^2$ in the entire range scanned.
\end{description}

In both cases above, $m_A,~m_{H^+}$ and the 4th generation fermion masses
can have values spanning over the entire scan ranges.
In Fig.~\ref{fig1} we plot the resulting distributions of the relevant parameter space
in the $S -T$, $\tan\beta - \sin\alpha$ and
$\Delta m_{\ell^\prime} - \Delta m_{q^\prime}$ planes, where
$\Delta m_{\ell^\prime} \equiv m_{\nu^\prime} - m_{\tau^\prime}$ and
$\Delta m_{q^\prime} \equiv m_{t^\prime} - m_{b^\prime}$. We
also show in Fig.~\ref{fig1} the resulting predicted 125 GeV Higgs signal
strengths for the two cases above, which, as can seen, have different characteristics.

We next discuss the compatibility of the above two 4G2HDM solutions with the recently observed
750 GeV $\gamma \gamma$ excess.

\section{The 4G2HDM and the 750 GeV $\gamma \gamma$ resonance \label{sec4}}

We search here for the portion of parameter space of the two 4G2HDM cases
found in the previous section, that survive once the 4G2HDM is also required to accommodate the 750 GeV
$\gamma \gamma$ excess, which is being
interpreted here as the decay of one or both of the heavy neutral Higgs (i.e.,
assumed to have masses $\sim 750$ GeV) $H \to \gamma \gamma$ and/or $A \to \gamma \gamma$.

Given the exploratory nature of our study,
we will simplify our analysis at this point, assuming that the scalar spectrum have the
characteristics
of the so-called decoupling limit (see e.g., \cite{decouplinglimit}). In particular,
we assume that it is
split into 2 typical scales:
$m_{light} \sim 125$ GeV, corresponding to the observed light Higgs
and $m_{heavy} \sim 750$ GeV around which the three heavy Higgs masses lie,
i.e., $m_H,m_A,m_{H^+} \sim 750$ GeV.
Even though we find a wider range of allowed masses for the non-resonant heavy scalar states
(i.e., for $m_A$ and $m_{H^+}$, see below)
that can accommodate the 750 GeV signal,
the choice $m_H,~m_A,m_{H^+} \sim 750$ GeV will suffice for
conveying our point: that the
750 GeV resonance in the $\gamma\gamma$ channel can be accommodated by
one of the heavy scalars of the 4G2HDM without any conflict with other existing relevant data.
Indeed, if this measurement will be eventually confirmed, then it will be instructive
to study the 4G2HDM within a wider range of the relevant parameter space.

We, thus, re-scan the 4G2HDM parameter space corresponding to two 4G2HDM cases found in
the previous section, where now $m_H$, $m_A$ and $m_{H^+}$ are
varied within a 30 GeV mass range around 750 GeV, i.e., $m_{H,A,H^+} \in 750 \pm 30$ GeV.
The scan is performed with the following additional ``filters"/requirements
(i.e., in addition to the requirement of compatibility with PEWD and with the measured
125 GeV Higgs signals, as outlined in the previous section):
\begin{itemize}
\item Reproducing the 750 GeV $\gamma \gamma$ excess within the range
$3 ~{\rm fb} < \sigma(pp \to H/A \to \gamma \gamma) < 13~{\rm fb}$. We find that
the (by far) dominant $H$ and/or $A$ production mechanism is the gluon-fusion
one $gg \to H/A$, so that all the relevant
cross-sections $\sigma(pp \to H/A \to f)$
are calculated in the narrow width approximation via:
\begin{eqnarray}
\sigma(pp \to H/A \to f) = \frac{C_{gg}}{s m_{H/A}}
\Gamma(H/A \to gg) BR(H/A\to f) ~,
\end{eqnarray}
where $\sqrt{s} =8$ or $13$ TeV and $C_{gg}$ is the gluon luminosity:
\begin{eqnarray}
C_{gg} = \frac{\pi^2}{8} \int_{m_{H/A}^2/s}^1 \frac{dx}{x} g(x)
g\left( \frac{m_{H/A}^2}{sx} \right) ~,
\end{eqnarray}
giving $C_{gg} \sim 2140(175)$ at $\sqrt{s} =13(8)$ TeV, see \cite{gluonPDF}.

\item The resonating scalar which produces the 750 GeV $\gamma \gamma$ excess is required
to have a width smaller than 45 GeV, i.e., $\Gamma_{H/A} < 45$ GeV.

\item We impose the existing experimental bounds
on the production and decays of the
heavy neutral scalars $H$ and $A$, as obtained at the 8 and 13 TeV LHC runs (in particular when applied to
$m_H,m_A \sim 750$ GeV) in all other channels which are
relevant to our study:
$pp \to W^+W^-,~ZZ,~t \bar t,~\tau \tau,~b \bar b,~hh,~hZ$.
In particular, we use the 95\% CL bounds in Table \ref{tab2}
quoted in \cite{1605.09401}.
\end{itemize}

\begin{table}[htb]
\begin{center}
\begin{tabular}{c||c|c|}
 final state & $\sigma$ at $\sqrt{s}=8$ TeV & $\sigma$ at $\sqrt{s}=13$ TeV \\
\hline \hline
 $pp \to H \to W^+ W^- $ & $< 40 $ fb & $< 300$ fb\\
\hline
 $pp \to H \to ZZ $ & $< 12 $ fb & $< 200$ fb\\
\hline
 $pp \to H \to hh $ & $< 39 $ fb & $< 120$ fb\\
\hline
$pp \to A \to hZ $ & $< 19 $ fb & $< 116$ fb\\
\hline
$pp \to H/A \to t \bar t $ & $< 450 $ fb &  \\
\hline
$pp \to H/A \to b \bar b $ & $< 1 $ pb &  \\
\hline
$pp \to H/A \to j j $ & $< 2.5 $ pb &  \\
\hline
$pp \to H/A \to \tau \tau $ & $< 12 $ fb & $< 60$ fb \\
\hline
\end{tabular}
\caption{Upper bounds at 95\% CL on $\sigma(pp \to S \to f)$
for various final states $f$, produced through a narrow resonance with
$m_S \sim 750$ GeV and $\Gamma_S/m_S \sim {\cal O}(10^{-2})$,
as applied to our scan with $S = H,A$.
The bound on $\sigma( pp \to H/A \to jj)$ is relevant for $j = {\rm gluon}$.
Table taken from \cite{1605.09401}.}
\label{tab2}
\end{center}
\end{table}

\begin{figure}
\begin{center}
\includegraphics[scale=0.37]{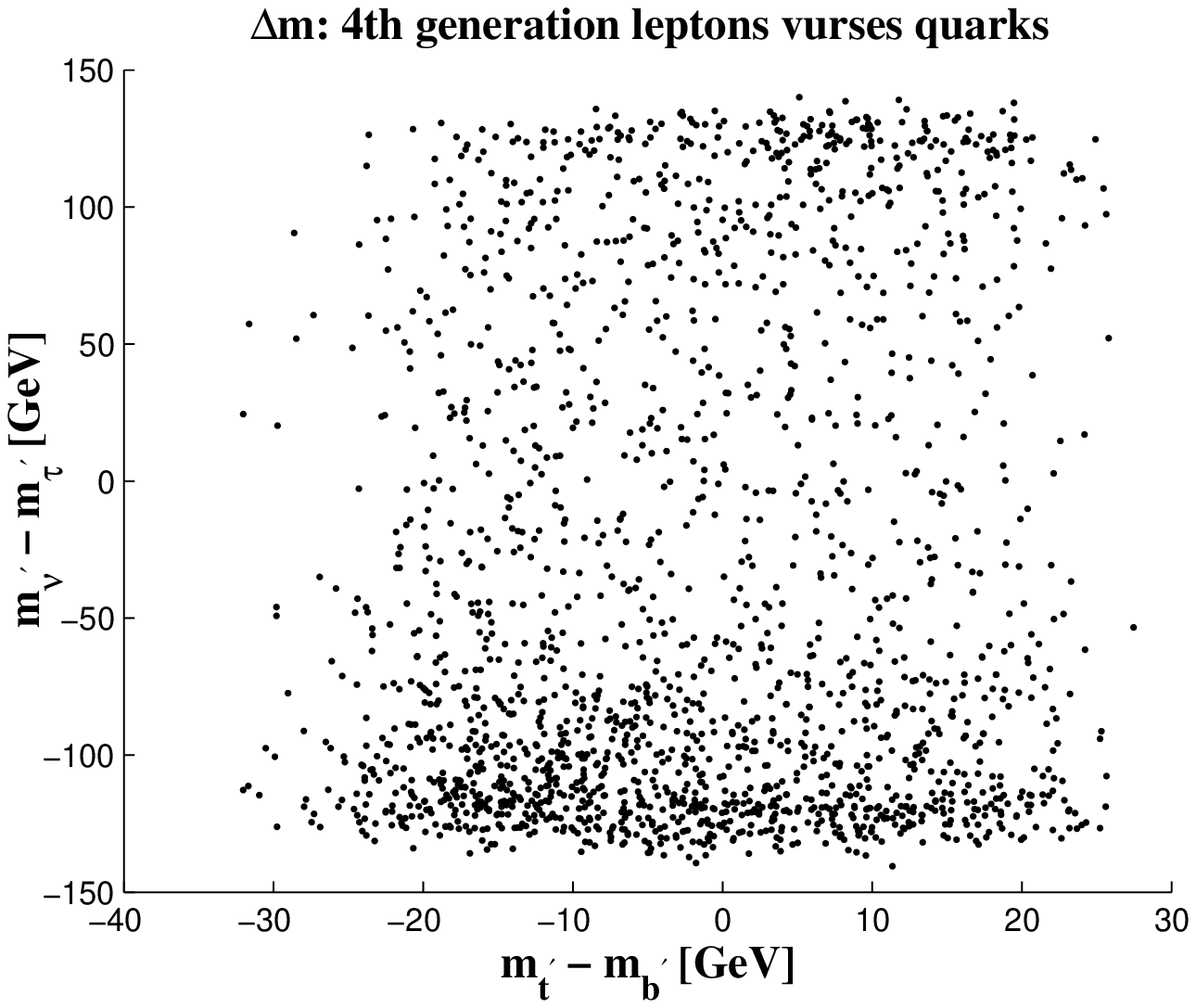}
\includegraphics[scale=0.37]{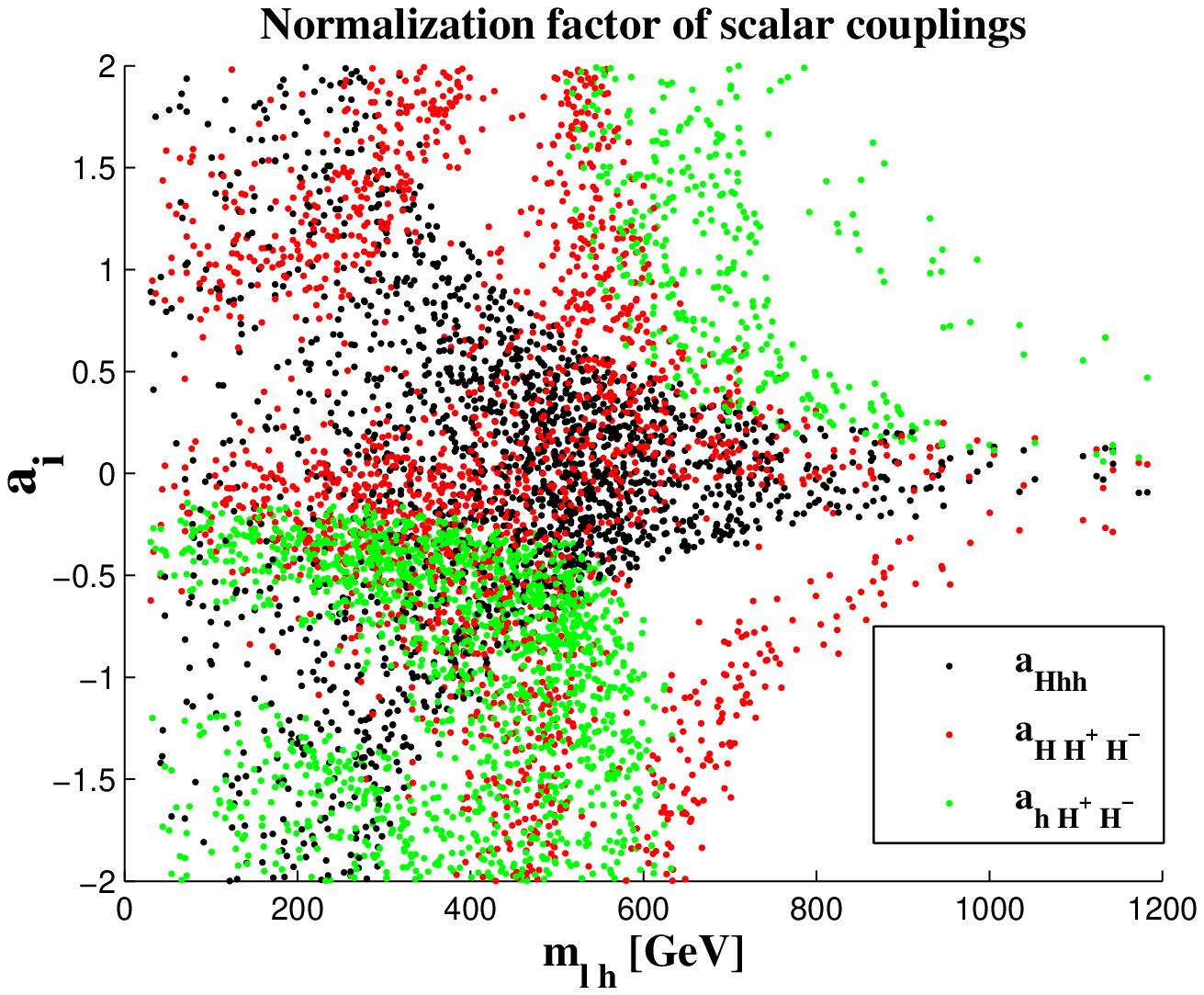}
\includegraphics[scale=0.37]{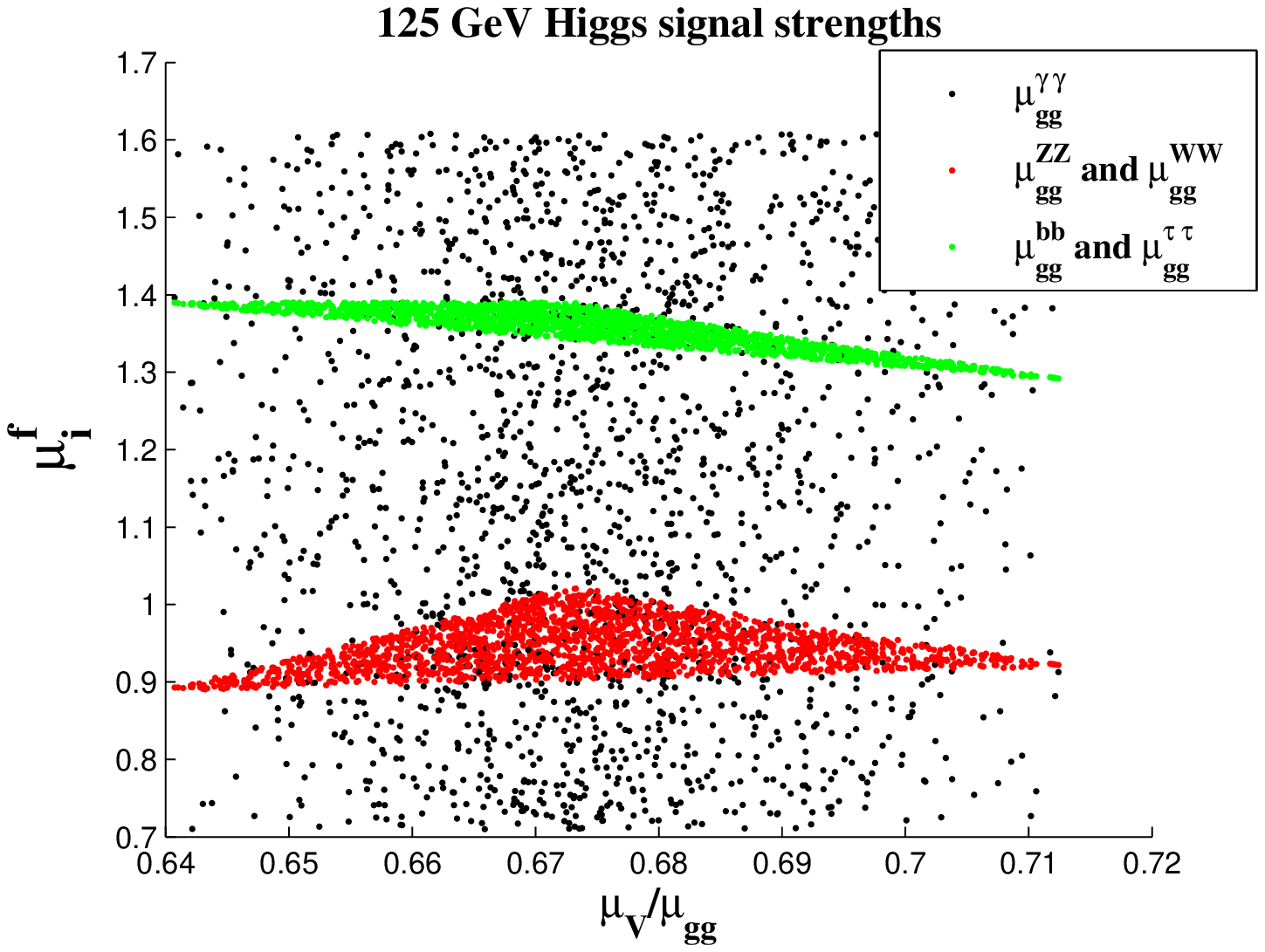}
\end{center}
\caption{Scatter plots of the 4G2HDM parameter space that is compatible
with the 125 GeV signals, with PEWD, with
$\sigma(pp \to H\to \gamma \gamma) = 3-13$ fb, with $\Gamma_H \leq 45$ GeV and
with all 8 and 13 TeV LHC bounds on the cross-section
$\sigma(pp \to H/A \to f)$ in all final states $f$ relevant to the $H$ and $A$ decays,
see Table \ref{tab2}. The scatter plots are given
for the mass splitting spectrum of the heavy fermions (left),
the correlation between the soft breaking mass
parameter $m_{\ell h}$ and the renormalization factor of the scalar couplings
$a_i = \bar\lambda_i/\lambda_i$, $i=Hhh,~ hH^+H^-,~HH^+H^-$ (middle), and the resulting
allowed ranges of the 125 GeV light Higgs signal strengths in all the measured channels (right).
\label{fig2}}
\end{figure}

Applying the above filters, we find that:
\begin{enumerate}
\item Only the CP-even scalar state $H$ (with $m_H =750$ GeV),
can accommodate the 750 GeV $\gamma \gamma$ resonance, since
$\sigma(pp \to A \to \gamma \gamma) \lsim {\cal O}(0.01)$ fb, which is 2-3 orders of
magnitudes smaller than the measured $\gamma \gamma$ excess, see also Table \ref{tab4}.
\item Only a ``shrinked" version of
the 4G2HDM case 1 survives out of the two cases that were found to be compatible with PEWD and
the 125 GeV light Higgs signals. In particular,
the surviving 4G2HDM models have (see Fig.~\ref{fig2}): $\tan\beta \leq 0.5$, $\alpha \to -\pi/2$
and $m_{\ell h} > 600$ GeV, having some correlation with the renormalization factors of the scalar couplings
$a_i = \bar\lambda_i/\lambda_i$, $i=Hhh,~ hH^+H^-,~HH^+H^-$.
\item The resulting heavy fermions mass ranges are narrowed to:
$350 ~ {\rm GeV} \lsim m_{t^\prime},m_{b^\prime} \lsim 390 ~ {\rm GeV}$, where the lower limit
is from direct searches (see section \ref{sec2}),
and $900 ~ {\rm GeV} \lsim m_{\nu^\prime},m_{\tau^\prime} \lsim 1200 ~ {\rm GeV}$,
where the upper limit is a rough estimate of the perturbativity bound on heavy chiral
leptons.
\end{enumerate}

In Fig.~\ref{fig2} we show three
scatter plots of the resulting 4G2HDM parameter space, corresponding to
the mass spectrum of the heavy fermions, the correlation between the soft breaking mass
parameter $m_{\ell h}$ and the renormalization factor of the scalar couplings
$a_i = \bar\lambda_i/\lambda_i$, $i=Hhh,~ hH^+H^-,~HH^+H^-$, and the resulting
allowed ranges of the 125 GeV light Higgs signal strengths in all the measured channels.
We see that, while $|m_{t^\prime} - m_{b^\prime}| \lsim 30$ GeV,
the mass splitting of the heavy leptons is typically $|m_{\nu^\prime} - m_{\tau^\prime}| \gsim m_W$.
We also see that smaller values of $m_{\ell h}$ typically require larger values of the
renormalization factors of the scalar vertices $a_i$, e.g., $a_{Hhh} \sim 1$
for $m_{\ell h} \sim 700$ GeV.

It is interesting to note that the resulting mass spectrum
of the heavy chiral quarks, which is required to accommodate the 750 GeV
$\gamma \gamma$ resonance, is rather narrow and
roughly centered around $m_H /2$, i.e.,
$m_{t^\prime},m_{b^\prime} \sim 350 - 390 ~ {\rm GeV}$.
This may hint back to the possibility that the
the heavy scalars are composites primarily
of the heavy chiral quarks, in which
case the 4G2HDM might indeed be interpreted
as a low energy effective framework for some TeV-scale strongly interacting
theory. Such an effective low energy 2HDM, with features similar
to the 4G2HDM discussed here,
was introduced in \cite{ourhybrid}, where it was shown that,
using the Nambu-Jona-Lasinio (NJL) mechanism \cite{Nambu}, it is possible to construct an
effective sub-TeV 2HDM hybrid framework, in which the 125 GeV light Higgs is mostly
a fundamental scalar, while the heavy Higgs states are components of
a composite field of the form
$\Phi_h \sim g_{t^\prime}^\star < \bar Q_L^{\prime c} (i \tau_2) t^{\prime c}_R > + g_{b^\prime} < \bar Q_L^\prime b^\prime_R >$, which is responsible
for EW symmetry breaking and for the dynamical mass generation of the heavy
quarks.$^{[4]}$\footnotetext[4]{Another interesting framework which entertains the idea that heavy chiral quarks
may form the 750 GeV composite was recently suggested in \cite{1602.05539}.}

\section{Phenomenology of the 4G2HDM \label{sec5}}

Inspired by the indications of the 750 GeV $\gamma \gamma$ resonance
and following the analysis of the previous section,
we briefly consider here
some of the distinct phenomenological consequences of the 4G2HDM
with characteristics similar to those
required to accommodate such a heavy scalar resonance.

In particular, we will assume below that $\tan\beta \sim 0.5$
and $\sin\alpha \sim -1$, in which case the light 125 GeV Higgs of the 4G2HDM, $h$, does not
couple to $f^\prime f^\prime$, while the heavy CP-even Higgs, $H$,
does not couple to a pair of SM fermions (see Eqs.~\ref{Sff1}-\ref{Sff2} and Table \ref{tab3}).
Also, the 4th generation heavy fermions are assumed to have masses in the ranges
$350 ~ {\rm GeV} \lsim m_{t^\prime},m_{b^\prime} \lsim 400 ~ {\rm GeV}$
and $900 ~ {\rm GeV} \lsim m_{\nu^\prime},m_{\tau^\prime} \lsim 1200 ~ {\rm GeV}$,
and the dominant decay channels of the heavy quarks are
$t^\prime \to uh$ ($u=u,c$) and $b^\prime \to dh$ ($d=d,s,b$), with
corresponding branching ratios $ \gsim 0.5$,
due to small off diagonal-entries $\Sigma^u_{4i}$ ($i=1,2$) and/or
$\Sigma^d_{4i}$ ($i=1,2,3$)
(see Table \ref{tab3} and discussion in section \ref{sec2}).

\begin{table}[htb]
\begin{center}
\begin{tabular}{|c||c|c|c|}
 & \multicolumn{3}{c|}{Yukawa couplings in the 4G2HDM with $\sin\alpha \sim -1$} \\
\hline
  &  $v \cdot y(\bar f f)$
  & $v \cdot y(\bar f^\prime f^\prime)$
 & $v \cdot y(\bar f_i f_{j})$ ($i, j = 1-4, i\neq j$)  \\
\hline \hline
 $h$ & $-\frac{m_f}{\cos\beta}$ & 0 &  $\frac{\Sigma^f_{ij}}{\cos\beta} (m_{f_i} R + m_{f_j} L)$ \\
\hline
 $H$ & 0 & $\frac{m_f}{\sin\beta}$ &  $ \frac{\Sigma^f_{ij} }{\sin\beta} (m_{f_i} R + m_{f_j} L)$ \\
\hline
 $A$ & $-i I_f m_f \tan\beta$  & $i I_f m_{f^\prime} \cot\beta$ &  $i I_f \frac{\Sigma^f_{ij}}{\sin\beta \cos\beta} (m_{f_i} R - m_{f_j} L)$ \\
  \hline
\end{tabular}
\caption{Yukawa couplings of the neutral Higgs particles in the 4G2HDM with $\sin\alpha \to -1$
and assuming $\Sigma^f_{ij} \ll \Sigma^f_{44} =1$ for
$ij \neq 44$, see section \ref{sec2}. In the first column
$f$ is a SM fermion of the 1st-3rd generations, while in the second column
$f^\prime$ stands for a 4th generation
fermion. In the 3rd column $f_i$ correspond to any fermion of the $i$th generation. Also, $I_f = 1(-1)$
for up(down) type fermions.}
\label{tab3}
\end{center}
\end{table}

In Table \ref{tab4} we list three benchmark points (BMP1,BMP2,BMP3)
which have some distinct characteristics and which are compatible with PEWD,
with the 125 GeV Higgs signals, with the 750 GeV $\gamma \gamma$ signal
and with the LHC bounds on all relevant 750 GeV Higgs resonance channels
$pp \to H/A \to f$ given in Table \ref{tab2}.
For definiteness, we have generated the benchmark points for the case of
$m_H = 750 $ GeV and $m_A,m_{H^+} \sim m_H \pm 50$ GeV, but the discussion below has a
more general scope, i.e., with regard to some of the possible
phenomenological signatures of the 4G2HDM associated with the TeV-scale
heavy scalars of the model and independent of whether the 750 GeV
$\gamma \gamma$ resonance is confirmed or not.
The three benchmark points include cases where the 750 GeV Higgs total width ranges from
a few GeV to $\sim 45$ GeV, having a resonance cross-section to $\gamma \gamma$ between 4-12 fb.
They also correspond to cases where $BR(H/A \to \bar q^\prime q^\prime) \sim 1$ and
$BR(H^+ \to \bar q^\prime q^\prime) \sim 1$.

\begin{table}[htb]
\begin{center}
\begin{tabular}{|c||c|c|c|}
\hline
  & BMP1 & BMP2 & BMP3  \\
\hline \hline
 $m_{t^\prime},m_{b^\prime}, m_{H^+}, m_A$ [GeV] & $352,382,709,780$ & $384,373,795,778$ & $368,369,691,731$   \\
\hline
 $H$ total width, $\Gamma_H$ [GeV] & 43 & 4 & 17   \\
\hline
 $\sigma(pp \to H/A \to \gamma \gamma)$ [fb] & $4/0.004$ & $12/0.006$ & $8/0.005$    \\
\hline
 $BR(H \to \bar t^\prime t^\prime, \bar b^\prime b^\prime, \bar t t, hh,gg)$ & $0.94,0,{\cal O}(10^{-6}),{\cal O}(10^{-4}),0.06$ &
 $0,0.36,0.04,0.007,0.64$& $0.47,0.38,{\cal O}(10^{-6}),{\cal O}(10^{-4}),0.14$  \\
\hline
$BR(A \to \bar t^\prime t^\prime, \bar b^\prime b^\prime, \bar t t, hZ,gg)$ & $0.61,0.33,0.015,0.04,{\cal O}(10^{-4})$ & $0.33,0.61,0.016,0.05,0.004$ & $0,0,0.29,0.71,{\cal O}(10^{-4})$   \\
\hline
 $\sigma(pp \to H \to \bar t^\prime t^\prime, \bar b^\prime b^\prime, \bar t t, hh)$ [fb] &
 $15000,0,0.015,5$
  & $0,4000,0.65,109$ & $7500,6000,0.02,9$  \\
\hline
 $\sigma(pp \to A \to \bar t^\prime t^\prime, \bar b^\prime b^\prime, \bar t t, hZ)$ [fb] &
 $160,87,19,52$
  & $90,164,21,57$
   & $0,0,38,91$   \\
\hline
 $BR(H^+ \to t^\prime \bar b^\prime, t \bar b ,W^+h)$ & $0,0.31,0.69$ & $0.9,0.03,0.07$ & $0,0.32,0.68$    \\
\hline
\end{tabular}
\caption{Benchmark points
with some distinct characteristics, which are consistent with PEWD,
with the 125 GeV Higgs signals, with the 750 GeV $\gamma \gamma$ signal
and with the LHC bounds on all relevant 750 GeV Higgs resonant channels
$pp \to H/A \to f$ given in Table \ref{tab2}.}
\label{tab4}
\end{center}
\end{table}

In particular, if $m_H,m_A > m_{q^\prime}/2$, then $H/A \to \bar q^\prime q^\prime$
is open and typically dominates, having a branching ratio of ${\cal O}(1)$ (see Table \ref{tab4}).
In that case, we find that within the 4G2HDM parameter space discussed here,
the corresponding resonance cross-sections
for $\bar q^\prime q^\prime$ production at the 13 TeV LHC are typically
$\sigma(pp \to H \to q^\prime q^\prime) \sim {\cal O}(10)$ [pb] and
$\sigma(pp \to A \to q^\prime q^\prime) \sim {\cal O}(0.1)$ [pb],
(both $H$ and $A$ produced through
gluon-fusion $gg \to H/A$), so that in the case of $H \to q^\prime q^\prime$ (see Table \ref{tab4}), this is
about an order of magnitude larger than the QCD (continuum) $\bar q^\prime q^\prime$ production rate.
Therefore, if the 750 GeV $\gamma \gamma$ resonance persists,
one should also expect an observable resonance signal at least in the
$H \to \bar q^\prime q^\prime$ channel.

Let us, therefore, briefly investigate the signal $H \to \bar q^\prime q^\prime$
under more general grounds, i.e., when $m_H > m_{q^\prime}/2$ but not necessarily
$m_H \sim 750$ GeV.
For example, in the case of $H \to \bar t^\prime t^\prime$,
the $t^\prime$ will further decay either via the FC channels $t^\prime \to uh$ ($u=u$ or $c$) or
via the 3-body decay $t^\prime \to b^\prime W \to dhW$ ($d=d,s,b$), where $b^\prime W$ are either off-shell
or on-shell (i.e., when $m_{t^\prime} > m_{b^\prime}+m_W$, see Fig.~\ref{fig1}).
If the former case (i.e., $t^\prime \to uh$) dominates, then the resulting resonance signal should be searched for
in $pp \to \bar t^\prime t^\prime \to (jh)_{t^\prime}(jh)_{t^\prime}$ ($j$ is a light jet),
while if the 3-body $t^\prime$ decay dominates then
$pp \to \bar t^\prime t^\prime \to (jhW^+)_{t^\prime}(jhW^-)_{t^\prime}$.
In either case, the SM-like light Higgs ($h$) further decays
into $b \bar b$ or $WW$ with SM rates, giving rise to resonance signatures
of the form $pp \to (nj+mb+\ell W)_H$, with
$(n,m,\ell)=(2,4,0),(2,0,4),(2,2,2),(2,4,2),(2,2,4),(2,0,6),(0,2,6),(0,4,4),(0,6,2)$
and with unique kinematic features
that distinguishes them from more conventional signatures.
Similar signals are also expected for $H \to \bar b^\prime b^\prime$.
We recognize that these type of signals are very challenging and may require new strategies,
in particular, for reconstructing the parent $q^\prime$'s in such a high jet-multiplicity environment.

The decay pattern of the charged Higgs may also change in the 4G2HDM,
in particular for the case when $m_{H^+} > m_{t^\prime} + m_{b^\prime}$,
for which the decay of  $H^+$ into a pair of heavy 4th generation fermions can dominate (see BMP1 in Table \ref{tab4}).
In particular, taking $m_{t^\prime} \sim m_{b^\prime} \equiv m_{q^\prime}$
and assuming that $H^+$ is sufficiently heavier than
$2m_{q^\prime}$, so that we can ignore corrections of
${\cal O}(4m_{q^\prime}^2/m_{H^+}^2)$ in the
phase-space factors, we have in the 4G2HDM:
\begin{eqnarray}
R_{t^\prime b^\prime/tb} &\equiv& \frac{\Gamma(H^+ \to t^\prime b^\prime)}{\Gamma(H^+ \to t b)}
\sim 2 \frac{m_{q^\prime}^2}{m_t^2} \cot^4\beta ~, \\
R_{t^\prime b^\prime/Wh} &\equiv& \frac{\Gamma(H^+ \to t^\prime b^\prime)}{\Gamma(H^+ \to Wh)}
\sim 12 \frac{m_{q^\prime}^2}{m_{H^+}^2} \left(\frac{\cot\beta}{\cos(\beta -\alpha)}\right)^2 ~.
\end{eqnarray}

Thus, for $\alpha \sim - \pi/2$, $\tan\beta \sim 0.5$ (i.e.,
$\cos(\beta - \alpha) \sim -0.45$), $m_{q^\prime} \sim 350$ GeV
(i.e., values of the 4G2HDM parameter space that can accommodate the 750 GeV $\gamma \gamma$ signal)
and taking $m_{H^+} \sim {\cal O}(1)$ TeV, we obtain: $R_{t^\prime b^\prime/tb} \sim {\cal O}(100)$
and $R_{t^\prime b^\prime/Wh} \sim {\cal O}(10)$, in which case
$BR(H^+ \to t^\prime b^\prime) \sim 1$ (e.g., as in the case of BMP2), leading to some interesting signatures
of the heavy charged Higgs at the LHC. In particular, the dominant production channels
of $H^+$ at the LHC are $gg/gb \to H^+ b \bar t, H^+ W^-/ H^+ \bar t$,
with a typical cross-section of $\sim 100$ fb when $\tan\beta \sim 1$ \cite{Hplusprod}.
The subsequent $H^+$ decay to a pair of 4th generation heavy fermions with
$BR(H^+ \to t^\prime \bar b^\prime) \sim 1$ will, thus, lead
to new $H^+$ signals, e.g.,
$pp \to t (t^\prime b^\prime)_{H^+} \to (b W)_t (jh)_{t^\prime} (jh)_{b^\prime}$, again
with the typical 4G2HDM heavy fermion high jet-multiplicity
signatures of the form $pp \to nj+mb+\ell W$.
This is in contrast to ``standard" 2HDM frameworks where the heavy charged Higgs
will dominantly decay to $Wh$ and/or $tb$ (see BMP1 and BMP3), leading to a lower multiplicity
of jets in the final state.

As noted earlier, a wider range of solutions exist (which are not being discussed here) to all data
and filters mentioned above (i.e., including the 750 GeV $\gamma \gamma$ resonance),
in which lighter pseudoscalar
$A$ and charged Higgs $H^{+}$ are allowed, with masses as low as $300$ GeV.
In such 4G2HDM scenarios, the heavy 4th generation quarks (and leptons) can
have substantial decay rates in channels involving also the heavy Higgs
species, i.e., $t^\prime \to H^+ d, Au$ ($d=d,s,b$ and $u=u,c)$
and $b^\prime \to H^+ u, Ad$ ($d=d,s,b$ and $u=u,c)$,
followed by $H^+ \to W^+ h, t \bar b$ and $A \to hZ, t \bar t$. Indeed, such decay patterns
can also lead to some un-explored collider signatures of the 4G2HDM.
We leave the discussion of the phenomenology of such
wider range of 4G2HDM scenarios to a later work.

Finally, we wish to comment on the flavor violating structure of the 4G2HDM and its
compatibility with the recently reported indications of the LFV
decay of the 125 GeV light Higgs $h \to \tau \mu$ \cite{ATLAS2,CMS2}.
Writing the LFV couplings of $h$ in a general form:
\begin{eqnarray}
{\cal L}(h f_i f_j) = {\cal S}_{ij} +  {\cal P}_{ij} \gamma_5 ~,
\end{eqnarray}
one obtains:
\begin{eqnarray}
\Gamma(h \to \bar f_i f_j + \bar f_j f_i) = \frac{m_h}{4 \pi} \left(
|{\cal S}_{ij}|^2 + |{\cal P}_{ij}|^2 \right) ~.
\end{eqnarray}

In our 4G2HDM we have for the case of the LFV decay $h \to \tau \mu$
(neglecting terms of ${\cal O}(m_\mu/m_\tau)$, see Eq.~\ref{Sff1}):
\begin{eqnarray}
|{\cal S}_{\tau \mu}| = |{\cal P}_{\tau \mu}| \sim
\frac{g}{4} \frac{m_\tau}{m_W}  f(\beta,\alpha) \xi_{\tau \mu}~,
\end{eqnarray}
where we have defined
$\Sigma^\ell_{32} = \Sigma^\ell_{23} \equiv \xi_{\tau \mu}$ (see Eq.~\ref{sigma}) and:
\begin{eqnarray}
f(\beta,\alpha) = \frac{\cos(\beta - \alpha)}{s_\beta c_\beta} ~.
\end{eqnarray}

Requiring now that $BR(h \to \tau \mu) \lsim 1\%$ we find:
\begin{eqnarray}
| f(\beta,\alpha) \xi_{\tau \mu}| \sim {\cal O}(0.1)~.
\end{eqnarray}

Thus, since for the values of $\tan\beta$ and $\alpha$ that were found to be compatible with
all data considered in the previous sections, we find
$|f(\beta,\alpha)| \sim 1-5$, and specifically $f(\beta,\alpha) \sim 1$ for
$\alpha \to -\pi/2$ and $\tan\beta \sim 0.5$, as required in order
to accommodate the 750 GeV $\gamma \gamma$ resonance (see previous section),
the 4G2HDM with $|\xi_{\tau \mu}| \lsim 0.1$ can address the measured
$BR(h \to \tau \mu) \lsim 1\%$ if it persists.

\section{Summary \label{sec6}}

We have revisited a class of models beyond the SM, suggested by us a few years
ago in \cite{ourpaper1}, which put together an additional Higgs doublet with a
heavy chiral 4th generation quark and lepton doublet and which have several important
and attractive theoretical features.
In particular, we focused on the so-called 4G2HDM of type I (in \cite{ourpaper1}),
in which a discrete $Z_2$ symmetry couples the ``heavy" scalar
doublet only to the heavy 4th generation fermions and the
``light" one to the lighter SM fermions.

We have confronted this model with PEWD, with the measured 125 GeV light Higgs signals
and also studied its compatibility with the recent indication of a 750 GeV $\gamma \gamma$ resonance
and with the current LHC bounds on heavy scalar resonances in other relevant channels.
We found that the CP-even heavy Higgs state of the 4G2HDM with a mass $\sim 750$ GeV
can accommodate the measured $750$ GeV excess for a rather unique
choice of the parameter space: $\tan\beta \sim 0.5$, $\alpha \sim -\pi/2$ (the Higgs mixing angle)
and with heavy chiral fermion masses
$m_{t^\prime,b^\prime} \lsim 400$ GeV and $m_{\nu^\prime,\tau^\prime} \gsim 900$ GeV.

We have shown that the heavy chiral quarks (and leptons) of the 4G2HDM
may have FCNC decays into the light 125 GeV Higgs plus a light-quark jet, $q^\prime \to j h$,
with branching ratios of ${\cal O}(1)$, thus leading
to some un-explored signatures of $q^\prime \bar q^\prime$ production at the LHC
and, therefore, being consistent with the current direct bounds
on the masses of new heavy fermions.
Indeed, new and rich phenomenology in $q^\prime$ - heavy Higgs systems is expected,
including possible resonance production of $q^\prime q^\prime$ pairs via either
the heavy neutral or heavy charged Higgs particles of the 4G2HDM, which leads to
high jet-multiplicity signatures, with or without charged leptons, of the form
$\bar q^\prime q^\prime \to nj +  mb + \ell W$, with $n+m+\ell=6-8$
and unique kinematic features which are related to the resonating heavy scalar
and the decay pattern of the heavy quarks.
The reconstruction of the $q^\prime q^\prime$ pairs in such high jet-multiplicity signals is very challenging and
require more thought and possibly new search strategies.

We also show that the recent indication of a percent-level branching ratio
in the LFV decay of the 125 GeV Higgs $h \to \tau \mu$, if it persists, can be readily
addressed within the distinct flavor structure of the 4G2HDM.

\bigskip
\bigskip
\bigskip
\bigskip
\bigskip
\bigskip

{\bf Acknowledgments:} We thank Pier Paolo Giardino for useful conversations.
The work of AS was supported in part by the US DOE contract \#DE-SC0012704.

\pagebreak

\newpage

\end{document}